\documentstyle{lamuphys}
\makeatletter
\let\chapter\hid@chapter
\makeatother
\begin{document}
\pagenumbering{arabic}
\title{ Critical Properties of 1-D Spin $1/2$
Antiferromagnetic Heisenberg Model }
\author{J.F.\,Audet\inst{1}, A.\,Fledderjohann\inst{2},
C.\, Gerhardt\inst{2}, M.\,Karbach\inst{2},
H.\,Kr{\"o}ger\inst{1}\thanks{talk presented by H. Kr{\"o}ger},
K.H.\,M{\"u}tter\inst{2} and M.\,Schmidt\inst{2}}

\institute{ D{\'e}partement de Physique, Universit{\'e} Laval,
Qu{\'e}bec, Qu{\'e}bec, G1K 7P4, Canada \\
Email: hkroger@phy.ulaval.ca \\
\and
Physics Department, University of Wuppertal,
D-42097 Wuppertal, Germany }
%
%
\maketitle
\begin{abstract}
We discuss numerical results for the $1-D$ spin $1/2$ antiferromagnetic
Heisenberg model with next-to-nearest neighbour coupling and in the presence of
an uniform magnetic field. The model develops zero frequency excitations at
field dependent soft mode momenta. We compute critical quantities from finite
size dependence of static structure factors.
\end{abstract}

\section{Introduction}
The study of quantum spin systems in one space dimension (spin chain)
can be motivated as follows:
There are models which are exactly soluble, e.g., the Heisenberg model.
"Exact" numerical results can be obtained by diagonalization of the Hamiltonian
for reasonably large chains (up to approximately 50 spins).
These quite precise data allow to carry out a finite size analysis
and extrapolate to the thermodynamic limit. On the other hand, $1-D$ spin
chains have been analyzed by different techniques, like quantum Monte Carlo
methods,
coupled cluster method ($exp(S)$-method), renormalization group techniques
etc.,
and thus constitute a benchmark problem for comparison of various methods.
{}From the physics point of view, there are substances occuring in nature
having
predominantly the structure of a $1-D$ chain. Examples are
$Cu Cl_{2} 2NC_{5}H_{5}$ or $Cu (NH_{3})_{4} SO_{4} H_{2}O$
\cite{Mazz55,Duni57,Bern81}.
New neutron scattering experiments on the quasi 1-D antiferromagnet
$K Cu F_{3}$ are discussed by Tennant et al. \cite{Tenn95}. There are recent
investigations by Dender et al. \cite{Dend96} on Copper benzoate also being a
linear chain antiferromagnet.
The Heisenberg model has a relation to the Hubbard model. According to
Inui\cite{Inui88} the pair hopping term of the Hubbard model can be mapped onto
the next-to-nearest neighbour interaction of the antiferromagnetic Heisenberg
(AFH) model.
{}From the theoretical point of view, $1-D$ spin chains are interesting because
its critical properties are predicted by conformal symmetry under the
hypothesis that this symmetry holds.
Finally, $1-D$ spin chains in the presence of magnetic fields and next-to
nearest neighbour coupling display quite a rich phase structure.
It is the purpose of this work to consider $1-D$ spin chains and to present
numerical results from exact diagonalization, discuss
its finite-size analysis and critical behavior. We mainly consider static spin
structure factors. In particular we analyze
chains with next-to-nearest neighbour coupling allowing for frustration.

\section{Heisenberg model and its extensions}
Let us start by recalling the definition of various $1-D$ spin models
considered in the following:
The classical model, describing next-neighbour interaction is given by the
Ising model
\begin{equation}
H = J \sum_{i=1}^{N} S_{i}^{z} S_{i+1}^{z},
\end{equation}
where $S_{i}^{z}$ is the z-component of spin $i$, taking the values $-1,+1$.
The corresponding quantum model is the Heisenberg model,
\begin{equation}
H = J \sum_{i=1}^{N} \vec{S}_{i} \cdot \vec{S}_{i+1},
\end{equation}
where $\vec{S}=\vec{\sigma}/2$ and $\sigma_{a}$ being the Pauli-matrices
for the spin 1/2 case. This is considered as the standard model of magnetism.
In the case $J>0$ the model describes antiferromagnetism, while $J<0$
describes the ferromagnetic system. The following generalizations are of
physical interest:
The $XXZ$ model describes the Heisenberg model with an anisotropy in
z-direction,
\begin{equation}
H = J \sum_{i=1}^{N} S_{i}^{x} S_{i+1}^{x} + S_{i}^{y} S_{i+1}^{y} +
cos(\gamma) S_{i}^{z} S_{i+1}^{z}.
\label{eq:xxzmodel}
\end{equation}
The Heisenberg model exposed to an external (classical) magnetic field is given
by
\begin{equation}
H = J_{1} \sum_{i=1}^{N} \vec{S}_{i} \cdot \vec{S}_{i+1} +
J_{B} \sum_{i=1}^{N} \vec{B}_{i} \cdot \vec{S}_{i}.
\label{eq:extmagnfield}
\end{equation}
If one takes into account next-to-nearest neighbour coupling of spins, the
model becomes
\begin{equation}
H = J_{1} \sum_{i=1}^{N} \vec{S}_{i} \cdot \vec{S}_{i+1} +
J_{2} \sum_{i=1}^{N} \vec{S}_{i} \cdot \vec{S}_{i+2},
\;\;\; \alpha =J_{2}/J_{1}.
\label{eq:nextnearneighb}
\end{equation}
If the sign of $J_{1}$ and $J_{2}$ are such that one term favors parallel
spins, while the other favors antiparallel spins, the system becomes
frustrated.
A model of particular interest with coupling beyond next neighbours is the
Haldane-Shastry model \cite{Hald88,Shas88} (with periodic boundary conditions),
\begin{equation}
H = J \sum_{n=1}^{N-1} \sum_{m=1}^{N} \frac{1} {\sin^{2}(n\pi/N) }
\vec{S}_{m} \cdot \vec{S}_{m+n}.
\label{eq:HaldShasmodel}
\end{equation}
Below we will consider combinations of the above cases, e.g., the presence of a
magnetic field and next-to-nearest neighbour coupling.
In the following we will consider spin models with periodic boundary
conditions.

\subsection{Symmetries and integrability of Heisenberg model}
The isotropic Heisenberg model has a number of symmetries and corresponding
conserved quantities: There is translational symmetry. The translation operator
$T = \exp[iP]$ defines the conserved quantum number of momentum $P$. There is
also reflexion symmetry. The total spin $\vec{S} = \sum_{i=1} \vec{S}_{i}$ is
conserved. There are more conserved quantities like, e.g., $F_{3}=2i
\sum_{n=1}^{N} \epsilon_{ijk} S_{n-1}^{i} S_{n}^{j} S_{n+1}^{k}$
\cite{Lusc76,Fabr90}. The existence of conserved quantities can be traced back
to
the existence of a set of commuting transfer matrices $T(\lambda)$, satisfying
$[T(\lambda),T(\lambda')]=0$ and $[T(\lambda),H]=0$.
The existence of such transfer matrices for the $1-D$ Heisenberg model
can be proved from inverse scattering theory or via the Yang-Baxter equations
\cite{Baxt80}.
The set of transfer matrices implies the existence of infinitely many conserved
quantities like, e.g.,  the Hamiltonian
$\left. H \propto \frac{\partial}{\partial \lambda} \ln
T(\lambda)\right|_{\lambda=1/2} + const.$,
the total spin squared
$\left. \vec{S}^{2} \propto \frac{\partial^{2}}{\partial \lambda^{2}}
T(\lambda)\right|_{\lambda=\infty}$,
or
$F_{3} \propto \left. \frac{\partial^{2}}{\partial \lambda^{2}}
\ln T(\lambda)\right|_{\lambda=i/2} + const.$ \cite{Fabr90}.
Thus the Heisenberg model is an integrable system \cite{Baxt80}.
Analytical solutions have been obtained via the Bethe ansatz \cite{Beth31}.
Using the Bethe ansatz, Hulth\'en \cite{Hult38} has calculated the ground state
energy per site in the thermodynamic limit
$\epsilon_{0}=1/4-\ln 2$ (corresponding to $J=1/2$).
Analytical information on the $1-D$ model at the critical point can be obtained
by making the assumption of conformal invariance.
Cardy \cite{Card86,Card87} has shown that the spectrum of the transfer matrix
is determined by conformal invariance. This predicts, e.g., the leading term
of the finite size behavior of the ground state energy, or critical exponents
of the spin-spin correlation function.

Recently, much analytic progress has been made in the Heisenberg and related
spin models. R\"omer and Sutherland \cite{Rome93} have computed
the finite size dependence of the correlation function of the AFH model.
The dynamical correlation function of the Calogero-Sutherland model
has been computed by Ha \cite{Ha94} and the
single-particle Green's function of this model
has been determined by Zirnbauer and Haldane \cite{Zirn95}.
Essler et al. \cite{Essl96} have shown that correlation functions of the XXZ
antiferromagnet near the critical magnetic field can be expressed
in terms of solutions of Painlev\'e differential equations.

\subsection{Quantum numbers of lowest state for finite chains}
The quantum numbers of the lowest lying state for a finite spin chain have been
analyzed by Fabricius et al. \cite{Fabr91}.
For a given total spin $S$ and number of spins $N$ one finds the quantum number
of momentum $P$ of the lowest lying state to be as follows: For $N$ even holds
\begin{equation}
P(\mbox{mod} \; 2 \pi) = - \pi \times \left\{
              \begin{array}{c}
              S : N=4,8,12,\cdots \\
              S+1 : N=2,6,10,\cdots
              \end{array}
              \right.
\end{equation}
For $N$ odd, there is degeneration of momentum
\begin{eqnarray}
&&
P^{+}(\mbox{mod} \; 2 \pi) = - \pi \times \left\{
              \begin{array}{c}
              S(1+1/N) : N=3,7,11,\cdots \\
              S(1+1/N)+1 : N=5,9,13,\cdots
              \end{array}
              \right.
\nonumber \\
&&
P^{-}(\mbox{mod} \; 2 \pi) = - \pi \times \left\{
              \begin{array}{c}
              S(1-1/N)+1 : N=3,7,11,\cdots \\
              S(1-1/N) : N=5,9,13,\cdots
              \end{array}
              \right.
\end{eqnarray}
The absolute ground state corresponds to
\begin{eqnarray}
&&
P(S=0) = \left\{
              \begin{array}{c}
              0 : N=4,8,12,\cdots \\
              \pi : N=2,6,10,\cdots
              \end{array}
              \right.
\nonumber \\
&&
P^{\pm}(S=1/2) = \mp \pi/2 \times \left\{
              \begin{array}{c}
              (1+1/N) : N=3,7,11,\cdots \\
              (-1+1/N) : N=5,9,13,\cdots
              \end{array}
              \right.
\label{eq:quantumnumbers}
\end{eqnarray}
These are the quantum numbers corresponding to the isotropic Heisenberg model.
In the case of next-to-nearest neighbour coupling the situation becomes more
complicated. Then the quantum numbers of the lowest lying state will depend on
the model parameters, like $\alpha$.

\subsection{Magnetization curve and susceptibility}
Magnetization is defined by $\vec{M}=<\vec{S}>/N$, it is a function of the
external field $\vec{B}$. The functional dependence $M$ versus $B$,
the so-called magnetization curve is most easily obtained by computing the
ground state energy $\epsilon$ per spin in a given spin sector and using
\begin{equation}
\frac{ \partial \epsilon(M) } {\partial M} = B.
\end{equation}
This follows from
\begin{eqnarray}
H &=& H|_{B=0} + \vec{B} \cdot \vec{S},
\nonumber \\
E &=& E|_{B=0} + \vec{B} \cdot <\vec{S}>,
\nonumber \\
\epsilon &=& E/N = \epsilon|_{B=0} + \vec{B} \cdot \vec{M}.
\end{eqnarray}
The magnetization curve of the $AFH$ with anisotropy at temperature zero
is shown in Fig.[~\ref{fig:magnetization}]. One observes that when
magnetization reaches its maximal value $M=1/2$ (all spins aligned), then the
magnetic field $B$ becomes saturated.
\begin{figure}
\vspace{6.0cm}
\caption[ ]{ Magnetization curve computed from Bethe ansatz equations for chain
of length N=2048. $\gamma$ is the anisotropy parameter. Taken from
\cite{Karb94a}. }
\label{fig:magnetization}
\end{figure}
This can be understood as follows:
For very long chains, i.e., near the thermodynamic limit, and close to
saturation (nearly all spins aligned), then the change in energy when
flipping a single spin becomes independent of the magnetization $M$. Thus
$B=\partial \epsilon / \partial M = \mbox{const.}$ and $B$ does not increase
when $M$ increases.
The magnetization curve for the isotropic $AFH$ model is analytically known to
leading order in two limiting cases: (a)
Weak magnetic field and (b) near saturation. The case of weak magnetic field
has been computed by Griffith \cite{Griff64} as well as by
Yang and Yang \cite{Yang66} using the Bethe ansatz and solving integral
equations numerically. One obtains the expansion
\begin{equation}
M(B)= \frac{B}{\pi} \left[ 1 -a_{1} \frac{1}{\ln B} -a_{2}
\frac{ \ln |\ln B| }{\ln^{2} B} - a_{3} \frac{1}{\ln^{2}B} + \cdots \right].
\end{equation}
The lowest order is due to Griffith \cite{Griff64}. The coefficient $a_{1}=1/2$
has been computed by Babujian \cite{Babu83} and $a_{2}=1/4$ has been obtained
by Lee and Schlottmann \cite{Lee87}, while $a_{3}$ is unknown.
The magnetization curve near saturation is known due to Yang and Yang
\cite{Yang66},
\begin{equation}
M(B)= \frac{1}{2} - \frac{\sqrt{2}}{\pi} \sqrt{B_{c}-B} + O(B_{c}-B),
\label{eq:saturation}
\end{equation}
and $B_{c}=2$ for the isotropic case.
{}From the magnetization curve, one obtains the magnetic susceptibility
$\chi = \frac{ \partial M(B) } { \partial B }$.
The susceptibility corresponding to the magnetization curve of
Fig.[~\ref{fig:magnetization}] is shown in Fig.[~\ref{fig:susceptibility}].
\begin{figure}
\vspace{6.0cm}
\caption[ ]{ Susceptibility computed from Bethe ansatz equations for chain of
length N=2048. $\gamma$ is the anisotropy parameter. Taken from
\cite{Karb94a}. }
\label{fig:susceptibility}
\end{figure}
Information on antiferromagnetic ordering is encoded in the ground state
spin-spin correlation function $< 0 | \vec{S}(x) \cdot \vec{S}(0) | 0 >$.
For spin chains being isotropic with respect to coupling of x- and y-components
of spin, one distinguishes the transverse and the longitudinal correlation
function,
\begin{equation}
\omega_{j}(l,N) = 4 <0|S^{j}_{l+1} S^{j}_{1} |0>, \;\;\; j=1,2:
\mbox{transverse},
\; j=3: \mbox{longitudinal}.
\end{equation}
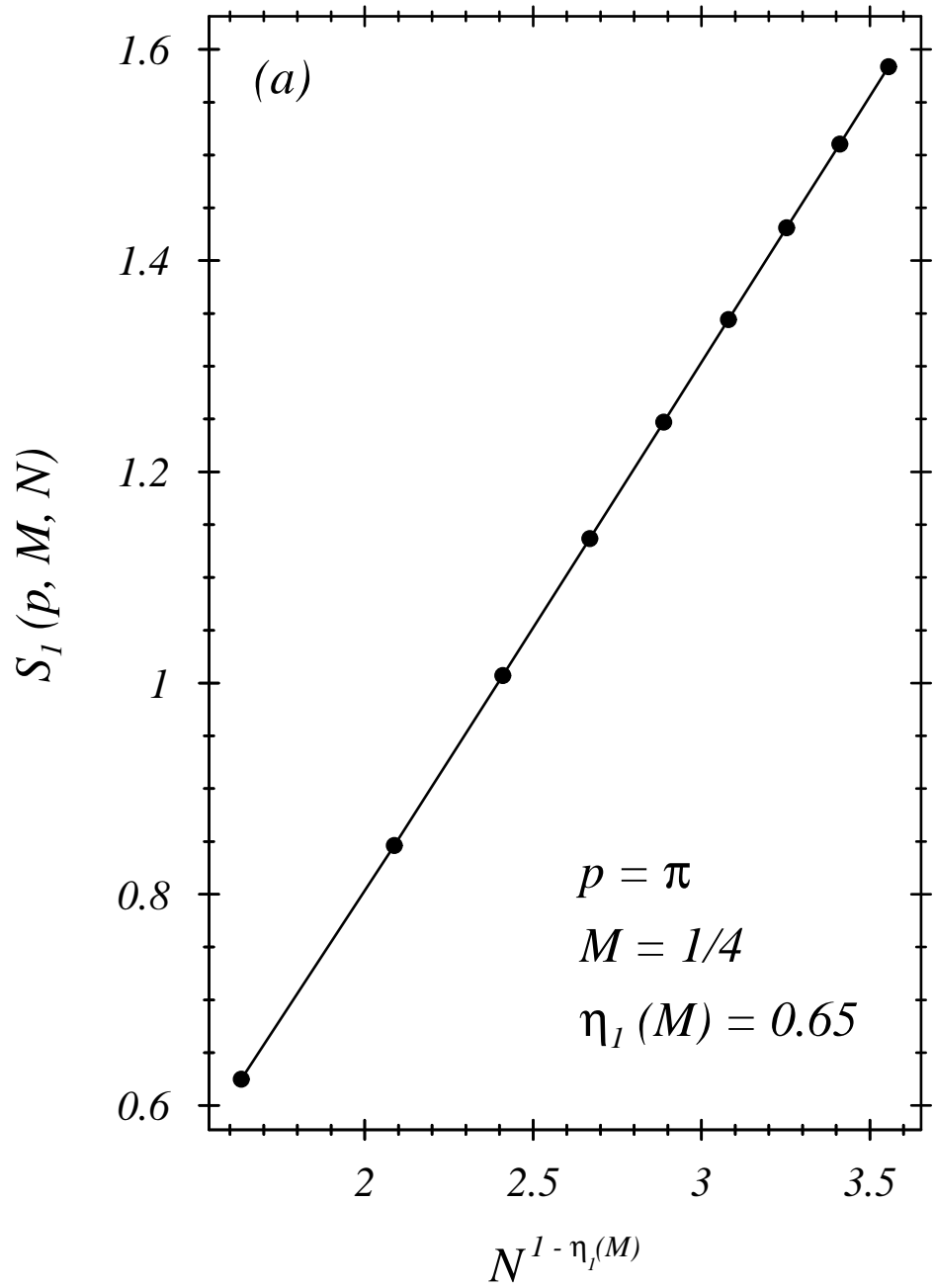
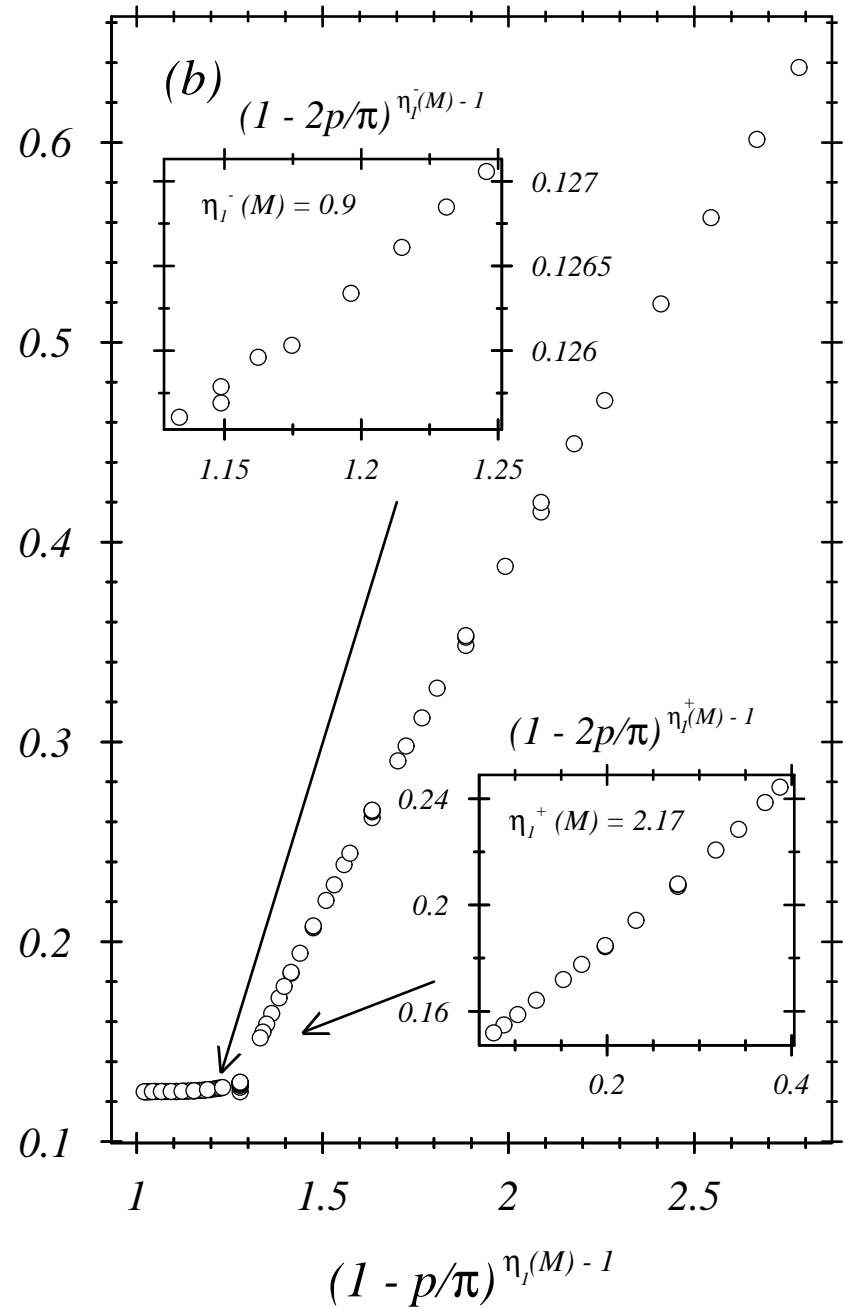
\begin{figure}
\vspace{13.0cm}
\caption[ ]{ Transverse (a) and longitudinal (b) structure factor for
$N=4, \cdots, 28$ and different anisotropy parameters $\gamma$.
Taken from \cite{Karb94a}. }
\label{fig:structurefactaniso}
\end{figure}
Luther and Peschel \cite{Luth75} have found the following behavior of the
correlation functions in the thermodynamic limit
\begin{equation}
\omega_{j}(l,N) \sim_{N \rightarrow \infty} \frac{(-1)^{l}}{l^{\eta_{j}}}.
\end{equation}
For the isotropic Heisenberg model the critical exponents $\eta_{j}$ obey
$\eta_{j}=1$ for $j=1,2,3$. This follows from conformal symmetry
\cite{Luth75,Bogo86}. Also based on conformal symmetry, Affleck et al.
\cite{Affl89} have later predicted for the isotropic Heisenberg model a
logarithmic correction to the long range behavior,
\begin{equation}
<S(x) S(0)> \sim
A \frac{(-1)^{x}}{x} (\ln x)^{\sigma}, \;\;\; \sigma=1/2.
\end{equation}
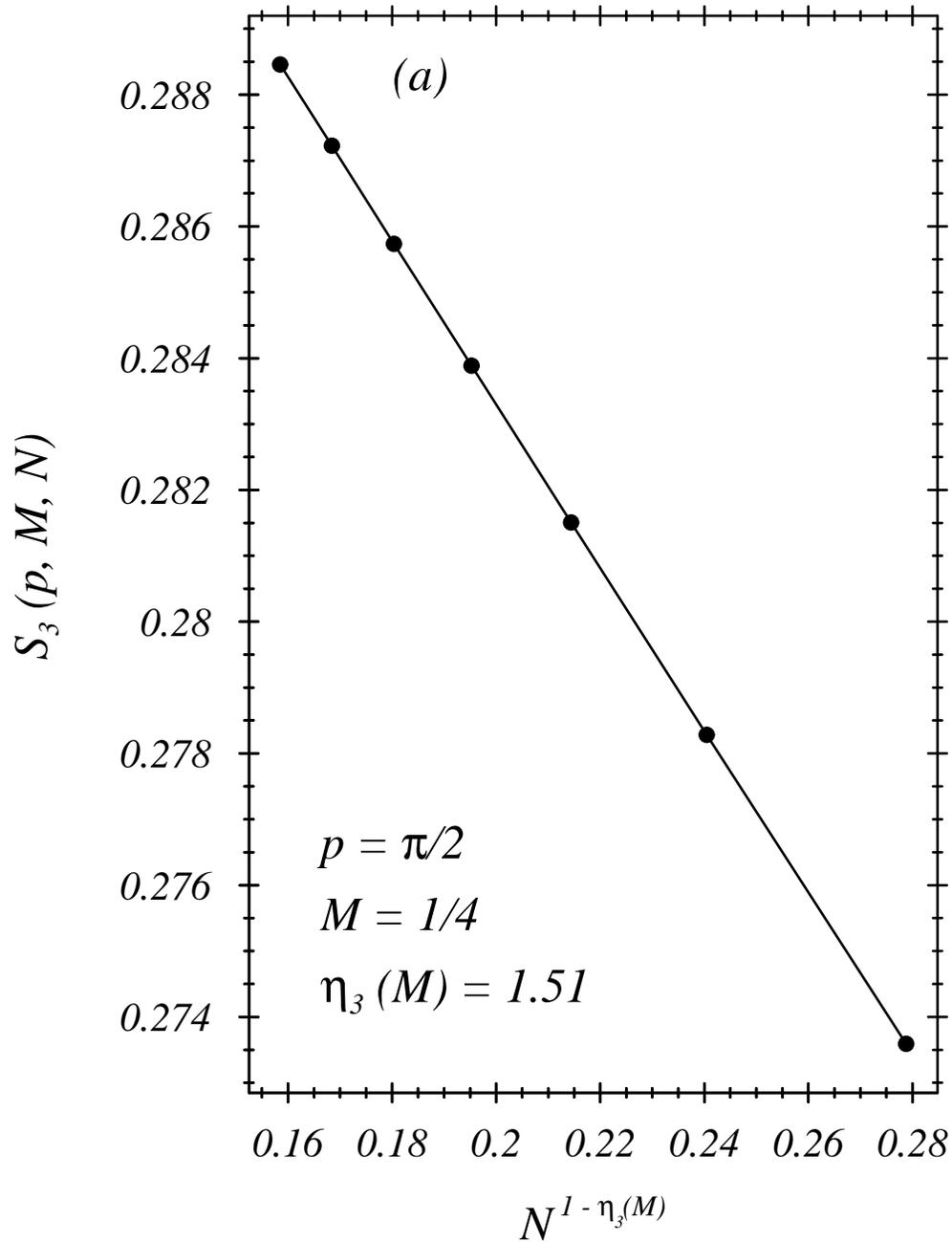
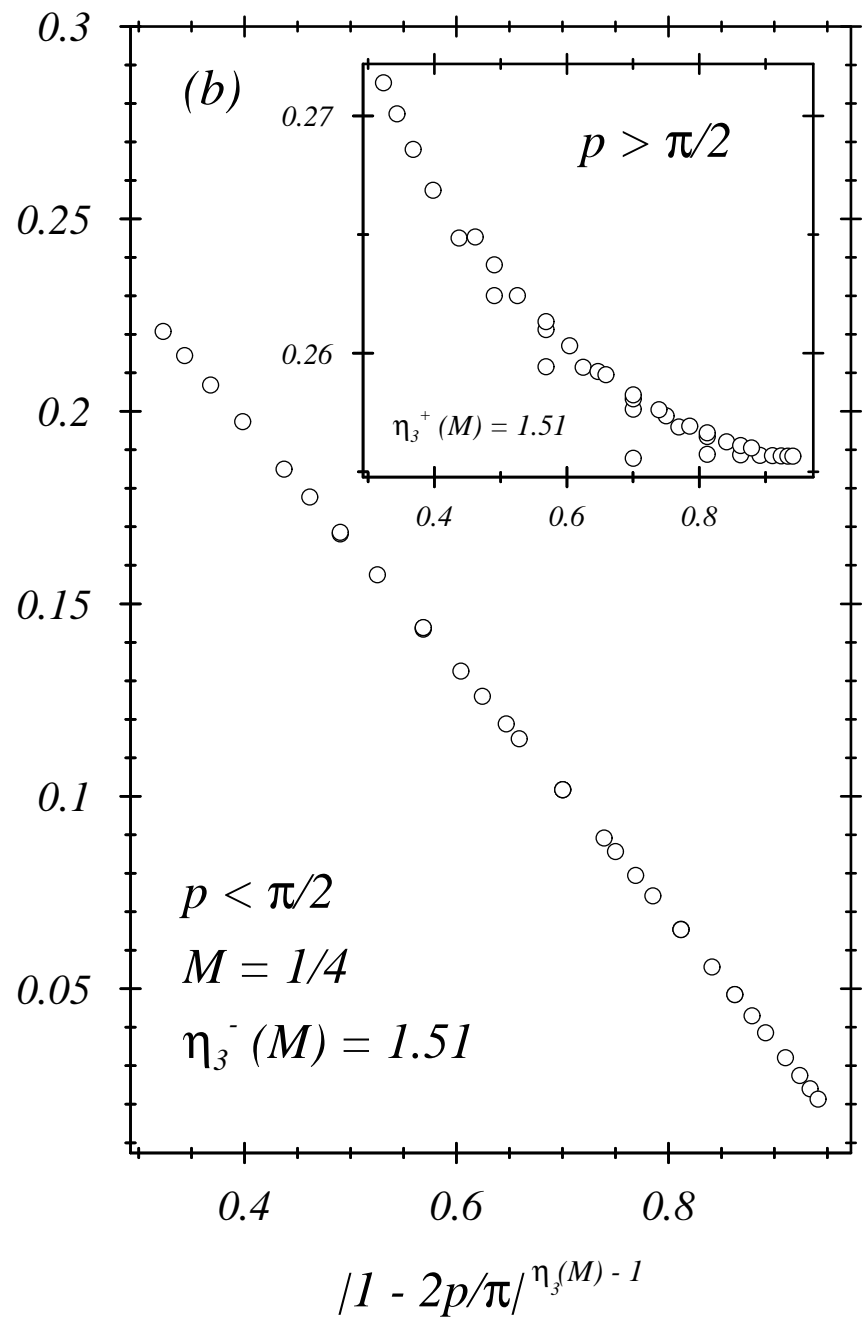
\begin{figure}
\vspace{6.0cm}
\caption[ ]{ Longitudinal structure factor versus scaling variable $L_{3}$,
Eq.(~\ref{eq:scalingvariable}), for $\gamma/\pi=0,0.1,0.2,\cdots,0.5$
and $N=4,6,\cdots,28$. }
\label{fig:scallongstruct}
\end{figure}

\section{Finite size scaling}
Critical behavior of a physical theory can exist only in the thermodynamic
limit ($N \rightarrow \infty$). Because very few models are soluble in the this
limit,
it is of great practical interest if physical information of the critical
system can be obtained from a finite number of spins (finite volume).
This has been answered by the theory of finite size scaling,
introduced by Fisher \cite{Fish71}.
It can be summarized as follows: Suppose a model becomes critical at a
temperature $T_{c}$. For the reduced temperature $t = (T-T_{c})/T_{c}$
this corresponds to $t_{c}=0$. Suppose there is an observable, which at the
critical point behaves like a power law
\begin{equation}
\Gamma(t) \longrightarrow_{t \rightarrow 0} \frac{1}{t^{\phi}},
\end{equation}
where $\phi$ is a critical exponent.
In a finite volume $N$ the observable is $\Gamma(t,N)$. The theory of
finite-size scaling is based on the following assumption relating the
observable in the thermodynamic limit to the finite volume,
\begin{equation}
\Gamma(t,N) \longrightarrow_{N \rightarrow \infty} \Gamma(t)
f(\frac{N}{\xi(t)}),
\end{equation}
where $\xi(t)$ is the correlation length in the thermodynamic limit
and $f(z)$ is some function of a single variable.
The correlation length has a critical behavior
\begin{equation}
\xi(t) \longrightarrow_{t \rightarrow 0} \frac{1}{t^{\nu}}.
\end{equation}
The finite-size scaling hypothesis can be written in an equivalent form,
originally proposed by Fisher \cite{Fish71},
\begin{equation}
\left.\Gamma(t,N) \longrightarrow_{N \rightarrow \infty} N^{\phi/\nu}
g(N t^{\nu})\right|_{\mbox{t=const.}}.
\end{equation}
According to a theorem by Mermin and Wagner \cite{Merm66} the Heisenberg model
in $D=1$ and $D=2$ dimensions
does not have long range order and hence does not have a 2nd order phase
transition for any finite temperature $T\neq 0$.
The critical exponent $\nu$ of the correlation length has been computed in the
thermodynamic limit for the $XXZ$ spin model at the critical point $T_{c}=0$
by Suzuki et al. \cite{Suzu92}, using the Bethe ansatz. The result is $\nu=1$.
\begin{figure}
\vspace{6.0cm}
\caption[ ]{ Finite-size dependence of $\Delta_{j}(\gamma,p,N)$,
Eq.(~\ref{eq:Delta}), at momentum $p=\pi/2$. }
\label{fig:finsizedelta}
\end{figure}

\section{Finite size analysis of spin-spin structure factors}
Although the Bethe ansatz in principle allows for an exact calculation
of the spin-spin correlation function, it is practically very difficult.
As an alternative many workers have computed numerically correlation functions
and structure factors.
The following standard methods have been applied successfully to spin systems:
exact diagonalization \cite{Bonn64},
coupled cluster - or $\exp(S)$ - method \cite{Bish95},
quantum Monte Carlo method \cite{Cepe95},
renormalization group techniques \cite{Whit93,Hall95}.
In the following we will focus on results obtained by exact diagonalization.
Due to limitations in computing time and storage, exact diagonalization is
possible only for small systems (for spin $1/2$ chain of $N$ spins in 1-D, the
dimension of Hilbert space is $2^{N}$). However, in the presence of a strong
magnetic field $B$ close to saturation nearly all spins are aligned. Setting up
a Hilbert space basis by counting configurations of non-aligned (reversed)
spins, the dimension will be much smaller. E.g., for $N=50$ spins in the sector
of magnetization $M=2/5$, the dimension of Hilbert space is $d=42376$. Thus
spin chains of
up to 56 spins have been diagonalized, using the Lan\c{c}zos method and binary
coding of spin configurations.
The advantage of exact diagonalization is the high numerical precision allowing
to carry out a finite size analysis and to extrapolate to
the thermodynamic limit.
\begin{figure}
\vspace{6.0cm}
\caption[ ]{ Thermodynamic limit of transverse (T) and longitudinal (L)
structure factor: $T1$, $L1$ finite-size analysis, $T2$, $L2$ conjecture of
Ref. \cite{Mull81}, $T3$ exact result of $XX$-model.
(a) $\gamma=\pi/2$, (b) $\gamma=0$. }
\label{fig:thermodynlim}
\end{figure}

\subsection{Structure factors in momentum space}
Physical information on magnetic ordering is encoded in the spin-spin
correlators. However, from the numerical point of view instead of analyzing the
long range behavior of the correlation function it is easier to
consider its Fourier transform, i.e., the spin structure factor in momentum
space \cite{Karb93}. The static spin structure factor is defined by
\begin{equation}
S_{j}(p,N) = \sum_{l=0}^{N-1} \omega_{j}(l,N) \exp[i p l], \;\;\; p=2\pi k/N,
\;\;\; k=0,\cdots, N-1.
\end{equation}
One has the following correspondence
\begin{eqnarray}
\mbox{x-space:} && <S(x) S(0) > \sim \frac{(-1)^{x}}{x},
\nonumber \\
\mbox{p-space:} && <S(p)> \sim \sum_{x} e^{i p x} \frac{(-1)^{x}}{x}
= \sum_{x} \frac{ e^{i (p+\pi) x} } {x} \;\;\; \mbox{divergent at} \; p=\pm\pi.
\end{eqnarray}
Thus the long distance behavior of the correlator manifests itself as
divergence in $p$-space. Such behavior can be analyzed numerically for finite
chains.
\begin{figure}
\vspace{6.0cm}
\caption[ ]{ Scaling function $g_{j}$ in the finite-size scaling ansatz of
Eq.(~\ref{eq:scalfctg}) for $z=2$. }
\label{fig:scalefctg}
\end{figure}

\subsection{Results for XXZ-model}
For the $XXZ$-model with anisotropy angle $\gamma$, Eq.(~\ref{eq:xxzmodel}),
the correlation function behaves as \cite{Luth75}
\begin{equation}
\omega_{j}(\gamma,l) \sim c(\gamma) \frac{(-1)^{l} } {l^{\eta_{j}(\gamma)} },
\;\;\;\ \eta_{1}(\gamma)= \eta_{3}^{-1}(\gamma) = 1 - \gamma/\pi, \;\;\;
0 \leq \gamma \leq \pi/2.
\end{equation}
Note that $\eta_{1}(\gamma) \cdot \eta_{3}(\gamma)=1$.
The corresponding structure factor behaves as
\begin{equation}
S_{j}(\gamma,p) \sim a_{j} + b_{j} \left[ 1 - p/\pi \right]^{\eta_{j}-1}.
\end{equation}
Because $\eta_{1}(\gamma)<1$ and $\eta_{3}(\gamma)>1$, at momentum $p=\pi$ the
transverse structure factor becomes divergent, while the longitudinal structure
factor remains finite. This behavior changes when $\gamma \rightarrow 0$
(isotropic limit). For finite systems the corresponding correlation function
and structure factor is given by
\begin{eqnarray}
\omega_{j}(\gamma,l) &\sim& \tilde{c}_{j}(\gamma) \frac{(-1)^{l} }
{l^{\eta_{j}(\gamma)} }
\exp \left[ l/\xi_{j}(\gamma,N) \right],
\nonumber \\
S_{j}(\gamma,p) &\sim& \tilde{a}_{j} + \tilde{b}_{j} \left[ (1 - p/\pi)^{2} +
1/\xi_{j}^{2}(\gamma,N) \right]^{\eta_{j}-1},
\end{eqnarray}
where $\xi_{j}(\gamma,N)$ is the correlation length in the finite system.
The correlation length has the following critical behavior when
$\cosh(\gamma')>1$ and $\gamma' \rightarrow 0$ \cite{John73}
\begin{equation}
\xi \longrightarrow \frac{1}{8} \exp \left[ \frac{\pi^{2}} {2 \gamma'} \right]
\;\;\;
\mbox{divergent at} \; \gamma'=0.
\end{equation}
The structure factor diverges at $p=\pi$ when $\gamma' \rightarrow 0$
\cite{Sing89}
\begin{equation}
S_{j}(\gamma',p=\pi,N=\infty) \longrightarrow_{\gamma' \rightarrow 0}
\frac{1}{\gamma'^{\lambda_{j}}}, \;\;\; \lambda_{1}=3/2, \lambda_{3}=2.
\end{equation}
The longitudinal and transverse structure factors for finite systems are shown
in Fig.[~\ref{fig:structurefactaniso}]
\begin{figure}
\vspace{6.0cm}
\caption[ ]{ Scaling behavior of transverse structure factor at $p=\pi$,
compared with finite-size scaling ansatz of
Eq.(~\ref{eq:finitesizeatpi}). }
\label{fig:scaletransstruct}
\end{figure}

Now let us look at the scaling properties of the structure factors.
One can ask the following questions:
(a) Does a scaling variable $L_{j}$ exist such that
$S_{j}(\gamma,p,N)=S_{j}(\gamma,L_{j}(\gamma,p,N))$?
(b) How does $S_{j}(\gamma=0,p,N=\infty)$ behave when
$p \rightarrow \pi$?
(c) What is the leading finite-size behavior of $S_{j}(\gamma,p,N)$
when $p < \pi$?
(d) How does $S_{j}(\gamma,p=\pi,N)$ behave when $N$ is large and
$\gamma \rightarrow 0$?
Karbach and M\"utter \cite{Karb93} have suggested to use $L_{3}(\gamma,p)$ as
scaling variable, given by
\begin{equation}
L_{3}(\gamma,p) = \frac{\eta_{3}(\gamma)}{\eta_{3}(\gamma) - 1}
\left[ 1 -(1 - p/\pi)^{\eta_{3}(\gamma)-1} \right],  \;\; p<\pi.
\label{eq:scalingvariable}
\end{equation}
Haldane and Shastry \cite{Hald88,Shas88} have shown that
\begin{equation}
\lim_{\gamma \rightarrow 0} L_{3}(\gamma,p) = - \ln(1 -p/\pi), \;\; p<\pi
\end{equation}
is the exact structure factor of the Haldane-Shastry model,
Eq.(~\ref{eq:HaldShasmodel}), with periodic boundary conditions.
Numerical studies \cite{Karb93,Karb94b} show the following scaling behavior.
Fig.[~\ref{fig:scallongstruct}] gives a plot of the longitudinal structure
factor versus the scaling variable $L_{3}$. The data points lie close to a line
of slope one when
$p < \pi$. Similar but less impressive scaling holds for the tranverse
component. Scaling deviations can be further analyzed by looking at
\begin{equation}
\Delta_{j}(\gamma,p,N)=S_{j}(\gamma,p,N)-L_{j}(\gamma,p).
\end{equation}
\begin{figure}
\vspace{6.0cm}
\caption[ ]{ Longitudinal structure factor versus momentum $p$ and
magnetization $M$ for $N=20,22,\cdots,28$. The ridge occurs at softmode
$p_{3}(M)$, Eq.(~\ref{eq:softmodemom}).  }
\label{fig:ridge}
\end{figure}
It is remarkable that at $\gamma=\pi/2$ ($XX$-model) no finite-size dependence
exists, $S_{3}(\gamma=\pi/2,p,N)=2p/\pi$.
For $p<\pi$ one makes the following finite size ansatz
\begin{equation}
\Delta_{j}(\gamma,p,N)=\Delta_{j}(\gamma,p,N=\infty) +
\frac{ c_{j}(\gamma,p) }{N^{2}}, \;\; p<\pi.
\label{eq:Delta}
\end{equation}
The finite size behavior is shown in
Fig.[~\ref{fig:finsizedelta}] and confirms the leading $1/N^{2}$ behavior when
approaching the thermodynamic limit.
The case $\gamma=\pi/2$ corresponds to the $XX$-model which is exactly soluble.
Fig.[~\ref{fig:thermodynlim}] shows the thermodynamic limit of the structure
factors.
\begin{figure}
\vspace{7.5cm}
\caption[ ]{ Transverse structure factor at $M=1/4$.
(a) Finite-size behavior at $p=\pi$, Eq.(~\ref{eq:S1finitesize}).
(b) Momentum dependence for $p \rightarrow \pi$, Eq.(~\ref{eq:S1momentum}).  }
\label{fig:S1}
\end{figure}
For $\gamma=\pi/2$ comparison with the exact result shows agreement.
The above analysis is not valid when approaching $p=\pi$.
The following finite-size scaling ansatz can be made in the combined limit
$N \rightarrow \infty$, $p \rightarrow \pi$,
\begin{equation}
\left. S_{j}(\gamma,p,N) \longrightarrow_{p \rightarrow \pi, N \rightarrow
\infty} \; S_{j}(\gamma,p,N=\infty) \;
g_{j}(\gamma,z)\right|_{z=(1-p/\pi)N=\mbox{const.}}
\label{eq:scalfctg}
\end{equation}
The scaling function $g_{j}$ is shown in Fig.[~\ref{fig:scalefctg}].
The transverse structure factor is singular at $p=\pi$. At the singularity the
following finite-size ansatz can be made
\begin{equation}
S_{1}(\gamma,p=\pi,N) \longrightarrow_{N \rightarrow \infty}
r_{1} \frac{\eta_{1}}{\eta_{1}-1} \left[ 1 - \sin(\eta_{1}\pi/2)
(a_{1}/N)^{\eta_{1}-1} \right].
\label{eq:finitesizeatpi}
\end{equation}
Numerical data compared with this ansatz are shown in
Fig.[~\ref{fig:scaletransstruct}].

\section{AFH model with next-to-nearest neighbour coupling and
external magnetic field}
Antiferromagnetic order is manifested in the Heisenberg model via singularities
of structure factors when $p \rightarrow \pi$, $N \rightarrow \infty$ and
$z=(1-p/\pi)N=\mbox{const}$.
Then the transverse structure factor becomes infinite, while the longitudinal
structure factor remains finite.
This antiferromagnetic order can be destructed (a) by switching on an external
uniform magnetic field $B$ or (b) by frustration (i.e., nearest neighbour and
next-to-nearest neighbour coupling do not favour the same alignment pattern).
Such will be examined in this section. As a result it turns out that there are
similarities between the anisotropic chain (XXZ) and the chain with
next-to-nearest neighbour coupling.
\begin{figure}
\vspace{7.5cm}
\caption[ ]{ Longitudinal structure factor at $M=1/4$.
(a) Finite-size behavior at softmode $p_{3}(M)$, Eq.(~\ref{eq:S3finitesize}).
(b) Momentum dependence for $p \rightarrow p_{3}(M)$,
Eq.(~\ref{eq:S3momentum}). }
\label{fig:S3}
\end{figure}

\subsection{Presence of external magnetic field}
When considering the Heisenberg Hamiltonian in the presence of a uniform
external magnetic field $\vec{B}$, Eq.(~\ref{eq:extmagnfield}), (with
$\vec{B}_{i}=\vec{B}=B\vec{e}_{z}$ and $J_{B}=1$),
one observes in the spectrum of the Hamiltonian the following property:
Zero-frequency modes $\omega(p_{soft})=0$ emerge
at non-zero "soft-mode" momenta $p_{soft}$. These modes show up in the
dynamical stucture factors $S_{1}(\omega,p)$ and $S_{3}(\omega,p)$
\cite{Fled96}.
The momenta depend on the field $B$, or the magnetization $M$, respectively.
One has
\begin{equation}
p_{1}(M)=2 \pi M, \;\;\; p_{3}(M)=\pi(1 - 2M).
\label{eq:softmodemom}
\end{equation}
The critical exponents $\eta_{j}$ are supposed to govern the long distance
behavior of the static structure factors and the infra-red behavior of the
dynamical structure factors. Now $\eta_{i}=\eta_{i}(M)$ become field dependent.
Provided that the low-lying excitation spectrum is governed by conformal
symmetry, one can make analytical predictions.
Numerical studies \cite{Karb95,Fled96}
show the following behavior.
In Fig.[~\ref{fig:ridge}] the longitudinal structure factor is plotted versus
momentum and magnetization. The results correspond to chains of
$N=20,22,\cdots,28$ spins.
The structure factor is a smooth function except for a ridge at the soft-mode.
One observes that the soft-mode moves with magnetization.
On the other hand, the transverse structure factor diverges at $p=\pi$ when $N
\rightarrow \infty$ like
\begin{equation}
S_{1}(p=\pi,M,N) \longrightarrow_{N \rightarrow \infty} N^{1 - \eta_{1}(M)}.
\label{eq:S1finitesize}
\end{equation}
For $M=1/4$ one finds, e.g., $\eta_{1}=0.65$, shown in Fig.[~\ref{fig:S1}a].
The exponent $\eta_{1}$ governs also the approach to the singularity at
$p=\pi$,
\begin{equation}
S_{1}(p,M,N=\infty) \longrightarrow_{p \rightarrow \pi}
(1-p/\pi)^{\eta_{1}(M)-1},
\label{eq:S1momentum}
\end{equation}
which is shown in Fig[~\ref{fig:S1}b].
When approaching the soft-mode momentum $p_{1}(M)$, one finds the behavior
\begin{equation}
S_{1}(p,M,N=\infty) \longrightarrow_{p \rightarrow p_{1}(M) \pm 0}
\left(1- p/p_{1}(M) \right)^{\eta_{1}^{\pm}(M)-1}.
\end{equation}
Numerical results indicate that $\eta_{1}^{-}(M)$ and $\eta_{1}^{+}(M)$
do not necessarily agree.
The corresponding behavior of the longitudinal stucture factor is as follows.
The finite size behavior at the soft-mode momentum $p_{3}(M)$ is given by
\begin{equation}
S_{3}(p_{3}(M),M,N) \longrightarrow_{N \rightarrow \infty} N^{1 - \eta_{3}(M)}.
\label{eq:S3finitesize}
\end{equation}
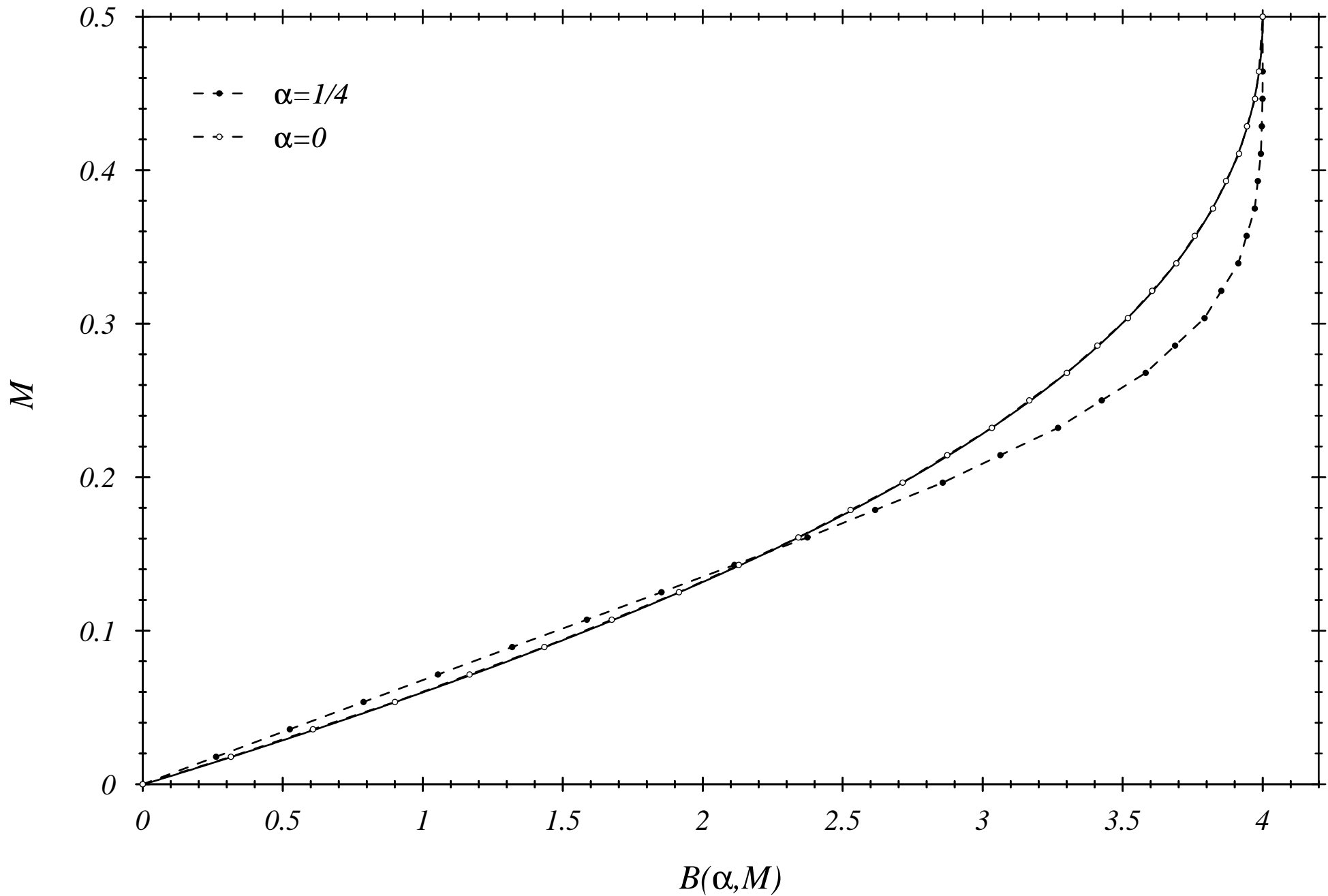
\begin{figure}
\vspace{6.0cm}
\caption[ ]{ Magnetization curve for $\alpha=0$ and $\alpha=1/4$.
The solid line is the Bethe ansatz solution with N=2048 spins. }
\label{fig:magnetizfrust}
\end{figure}
This is displayed in Fig.[~\ref{fig:S3}a].
Approaching the soft-mode from the left one has
\begin{equation}
S_{3}(p,M,N=\infty) \longrightarrow_{p \rightarrow p_{3}(M)-0}
\left(1- p/p_{3}(M) \right)^{\eta_{3}^{-}(M)-1},
\label{eq:S3momentum}
\end{equation}
shown in Fig.[~\ref{fig:S3}b].
There is indication for $\eta_{3}(M)=\eta_{3}^{-}(M)$. However, if
$\eta_{3}^{+}(M)$ and $\eta_{3}^{-}(M)$ agree is an open question.

\subsection{Next-to-nearest neighbour coupling}
Now let us consider the model Hamiltonian including next-to-nearest neighbour
coupling and the presence of a magnetic field, given by a combination of
Eqs.(~\ref{eq:nextnearneighb},~\ref{eq:extmagnfield}).
This model has a rich phase structure, depending on the parameters
$J_{1}$, $J_{2}$ and $B$. There is a ferromagnetic phase, an antiferromagnetic
phase and a frustrated phase \cite{Farn94}.
Haldane \cite{Hald82} has shown that within the antiferromagnetic phase
there is a phase transition at $\alpha_{c}$ between a spin-fluid phase
($\alpha < \alpha_{c}$) and a dimer phase ($\alpha > \alpha_{c}$). The
transition point has been determined to be $\alpha_{c}=0.24$ \cite{Okam92}.
Related to this is Gluzman's \cite{Gluz94}
prediction of the occurence of a plateau in the magnetization curve.
Moreover, it is remarkable that
the model is exactly soluble at $\alpha=1/2$, $B=0$,
called Majumdar-Ghosh point \cite{Maju69}.
\begin{figure}
\vspace{12.0cm}
\caption[ ]{ Transverse structure factor (a)
and longitudinal structure factor (b)
versus momentum $p$ for fixed magnetization $M=1/4$
but different values of $\alpha$. }
\label{fig:fixedmag}
\end{figure}
One can ask: Where are the observable differences in the model
with and without next-to-nearest neighbour coupling, i.e., $\alpha \neq 0$ and
$\alpha=0$?
Numerical studies by Schmidt et al. \cite{Schm96}
and Gerhardt et al. \cite{Gerh96}
give the following information.
Firstly, one observes in the magnetization curve near saturation the following
characteristic behavior,
\begin{equation}
M(\alpha=1/4,B) \longrightarrow_{B \rightarrow B_{s}} 1/2 - \mbox{const.}
\times
(B_{s} -B)^{1/4},
\end{equation}
in contrast to the square root behavior for $\alpha=0$,
given by Eq.(~\ref{eq:saturation}).
This is shown in Fig.[~\ref{fig:magnetizfrust}].
In the presence of an external magnetic field
and next-to-nearest neighbour coupling, the structure factors develop
singularities at soft-mode momenta, given by Eq.(~\ref{eq:softmodemom}).
Now the strength of the singularity will depend on the the frustration
parameter $\alpha$. The transverse structure factor and the longitudinal
structure factor
for fixed values of $\alpha$ but different values of magnetization $M$ are
shown in Fig.[~\ref{fig:fixedmag}a] and [~\ref{fig:fixedmag}b], respectively.
Fig.[~\ref{fig:fixedalp}] displays the structure factors
for fixed $M$ but different values of $\alpha$.
The critical exponents $\eta_{i}$ will now depend on $M$ and $\alpha$.
The finite-size dependence of the transverse structure factor at $p=\pi$ is
described by
\begin{equation}
S_{1}(\alpha,p=\pi,M,N) \longrightarrow_{N \rightarrow \infty}
A_{1}(\alpha,M) + B_{1}(\alpha,M) N^{1-\eta_{1}(\alpha,M)}.
\label{eq:S1frust}
\end{equation}
\begin{figure}
\vspace{12.0cm}
\caption[ ]{ Transverse structure factor (a) and longitudinal structure factor
(b) versus momentum $p$ for fixed $\alpha=1/4$ but different values of $M$. }
\label{fig:fixedalp}
\end{figure}
E.g., one finds $\eta_{1}(\alpha=0,M=1/4)=0.65$, while
$\eta_{1}(\alpha=1/4,M=1/4)=1.16$.
Similarly, the longitudinal structure factor at the soft-mode $p_{3}(M)$
behaves as
\begin{equation}
S_{3}(\alpha,p=p_{3}(M),M,N) \longrightarrow_{N \rightarrow \infty}
A_{3}(\alpha,M) + B_{3}(\alpha,M) N^{1-\eta_{3}(\alpha,M)}.
\label{eq:S3frust}
\end{equation}
E.g., one finds $\eta_{3}(\alpha=0,M=1/4)=1.5$, while
$\eta_{3}(\alpha=1/4,M=1/4)=0.84$.
Those critical exponents extracted from
Eqs.(~\ref{eq:S1frust},~\ref{eq:S3frust}) are given in Tab.[~\ref{tab:eta}] as
a function of $\alpha$.
Based on the assumption that conformal symmetry holds, one would expect
$\eta_{1} \cdot \eta_{3} = 1$.
The numerical data for most values of $\alpha$ deviate from the expected result
by a few percent. However, special attention should be paid to the region
$\alpha \sim -0.2$ and the region $\alpha \ge 0.4$.
For $\alpha =-0.2$ Eq.(~\ref{eq:S3frust}) does not yield a value of $\eta_{3}$.
It turns out that the finite-size depence is very weak in this region
($B_{3}(\alpha,M)$ is very small) and may even vanish for some value of
$\alpha$.
Secondly, for $\alpha \ge 0.4$, $\eta_{1} \cdot \eta_{3}$ deviates
substantially from the value one. The reason is most likely the following:
The structure factors have been computed in the sector corresponding to quantum
numbers of the Heisenberg model, Eq.(~\ref{eq:quantumnumbers}).
For $\alpha \ge 0.4$, those are presumably no longer the quantum numbers of the
ground state, hence the critical exponents $\eta_{i}$ do not correspond to the
ground state but to some excited state.
\begin{table}
\caption[ ]{Critical exponents $\eta_{i}$ for $M=1/4$ versus $\alpha$ }
\begin{center}
\begin{tabular}{|c|c|c|c|c|c|c|c|c|c|c|c|c|c} \hline\hline
$\alpha$                  & -0.5 & -0.4 & -0.3 & -0.2 & -0.1 & 0    & 0.1  &
0.2  & 0.25 & 0.3  & 0.4  & 0.5  \\ \hline
$\eta_{1}$                & 0.39 & 0.42 & 0.45 & 0.50 & 0.56 & 0.65 & 0.78 &
1.00 & 1.16 & 1.36 & 2.74 & 3.53 \\ \hline
$\eta_{3}$                & 2.52 & 2.45 & 2.34 &  ?   & 1.66 & 1.49 & 1.26 &
0.98 & 0.84 & 0.70 & 0.47 & 0.32 \\ \hline
$\eta_{1} \cdot \eta_{3}$ & 0.98 & 1.03 & 1.06 &  ?   & 0.93 & 0.97 & 0.98 &
0.98 & 0.97 & 0.94 & 1.28 & 1.13 \\ \hline
\end{tabular}
\label{tab:eta}
\end{center}
\end{table}

\section{Summary and outlook}
The 1-D spin 1/2 AFH model in the presence of an external magnetic field and
with next-to-nearest neighbour coupling can be solved "exactly"
by diagonalization (via Lan\c{c}zos) of chains of up to 50 spins (depending on
$M$). For the ground state properties of the model
this yields data with high numerical precision
allowing for finite-size scaling analysis and extraction of critical exponents.
Here we have concentrated on static spin structure factors.
One finds a different behavior for $\alpha < \alpha_{c}$ and
$\alpha > \alpha_{c}$. The difference is observed in the strength of the
singularities of structure factors and in the magnetization curve near
saturation.
Much interesting physics is to be explored when going into D=2,3 dimensions
(e.g., prediction of plateaus in the magnetization curve).
Exact diagonalization is presently limited in D=2 to $7 \times 7$ spins
(with helical boundary conditions). To analyze critical behavior in
D=2, new methods yielding numerically precise results for a larger number of
spins are needed.

\begin{flushleft}
{\bf Acknowledgement}
\end{flushleft}
One of the authors (H.K.) is grateful for support by NSERC Canada.

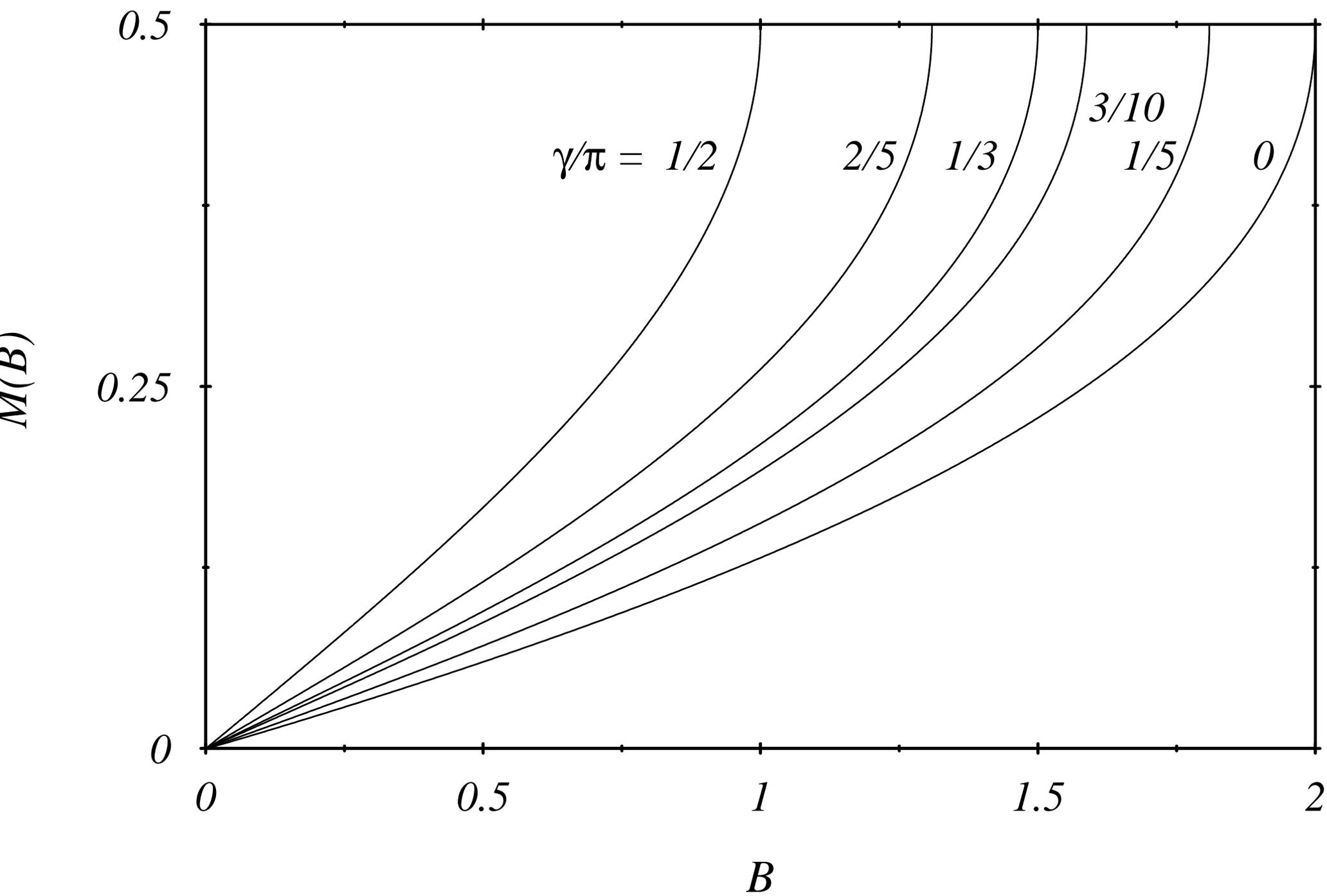
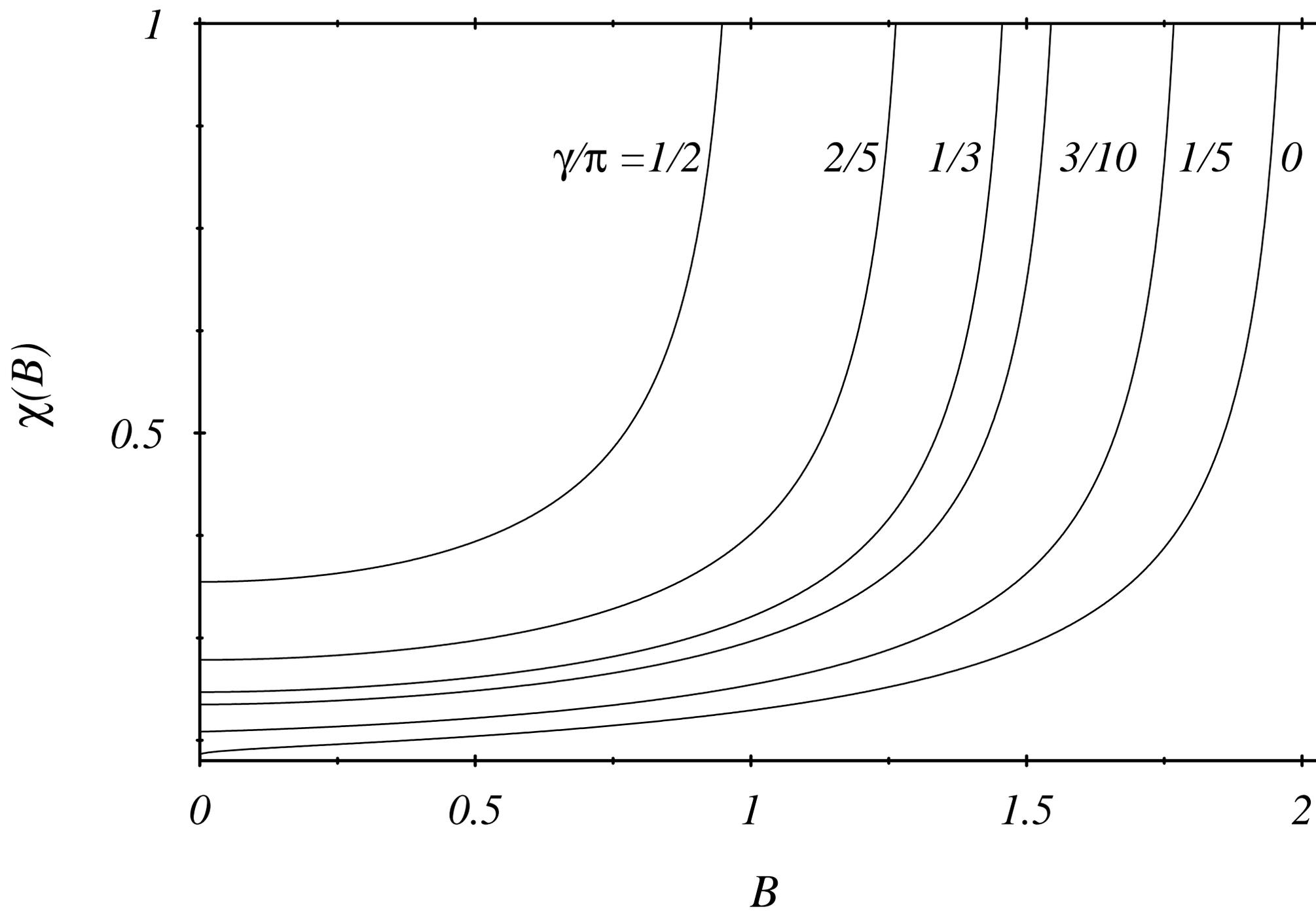
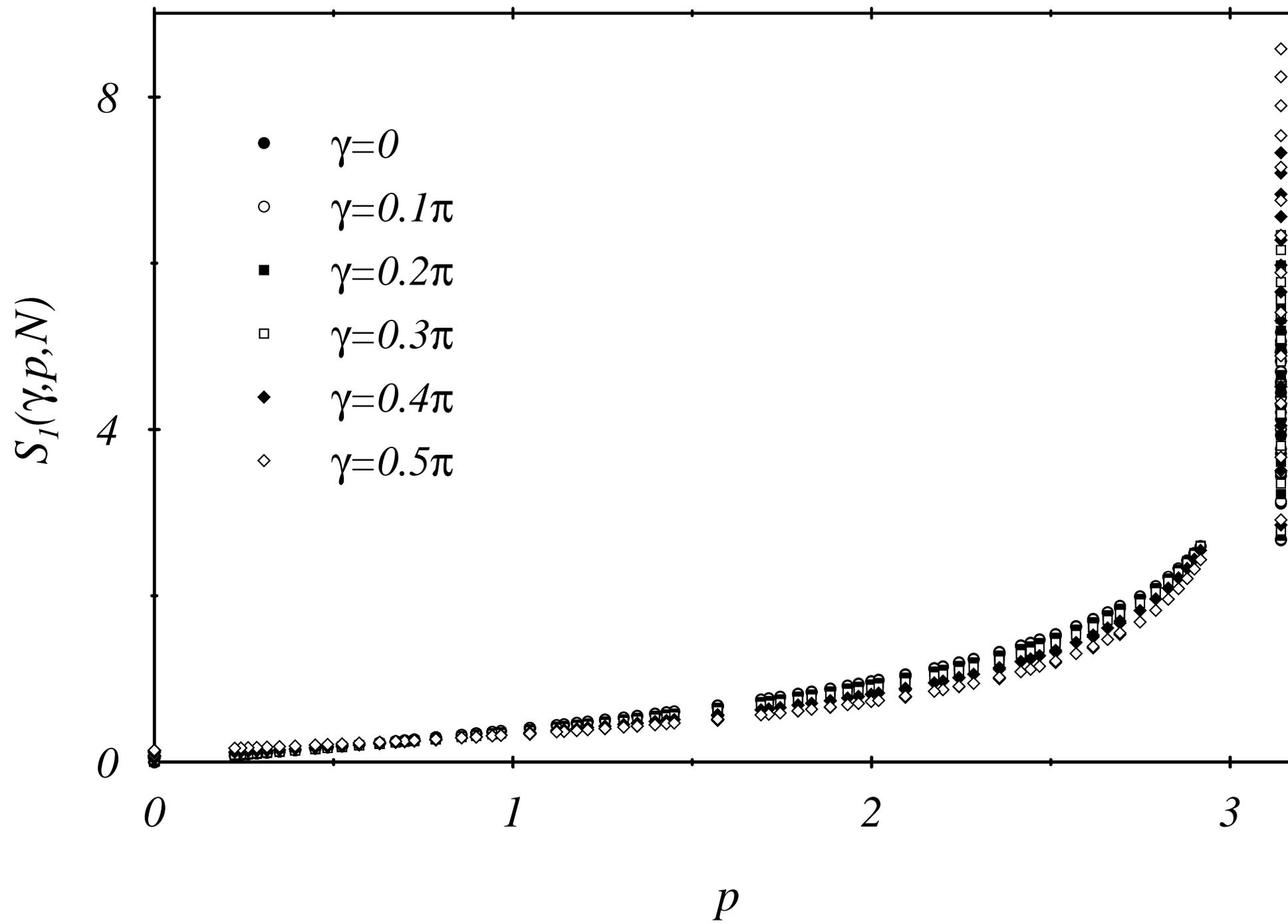
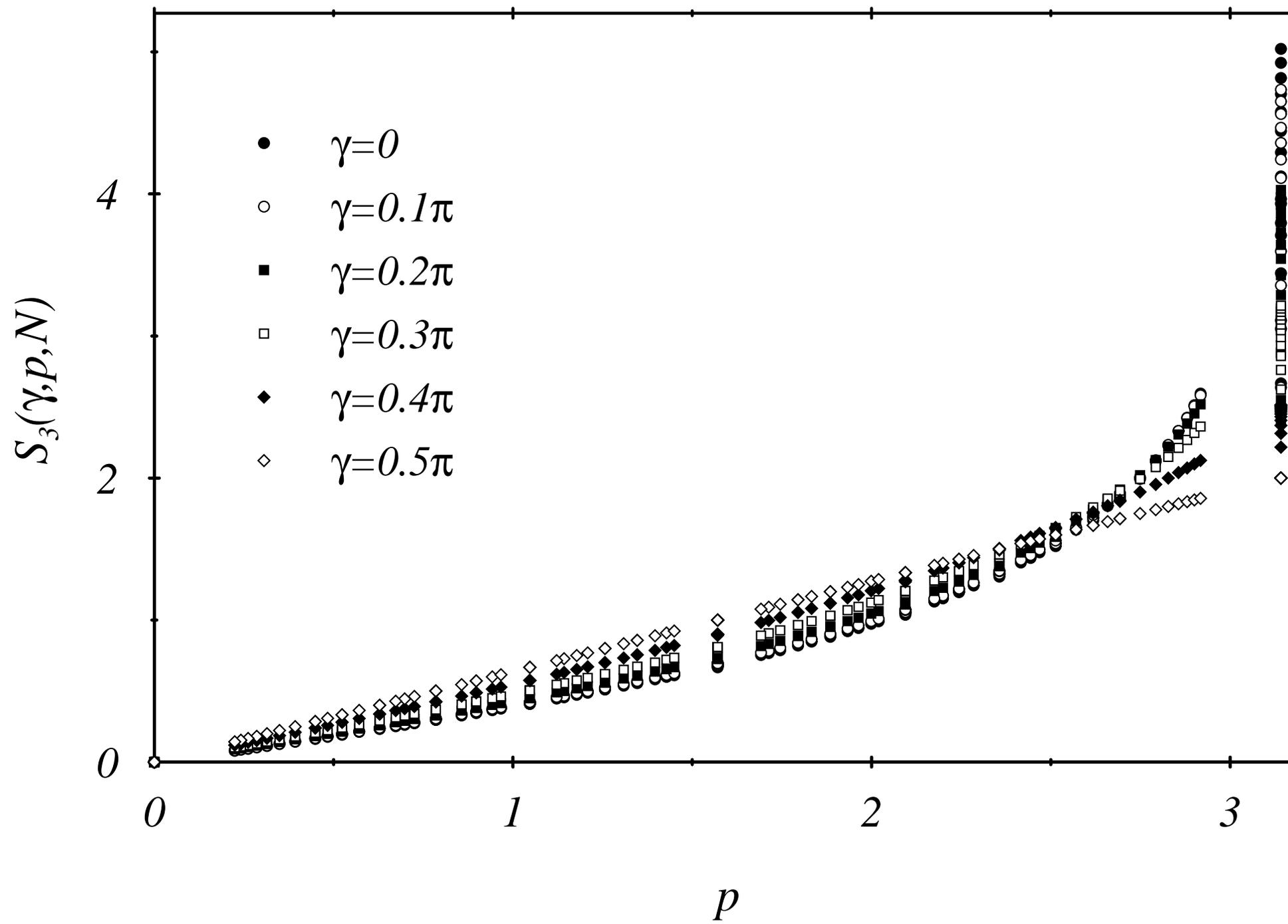
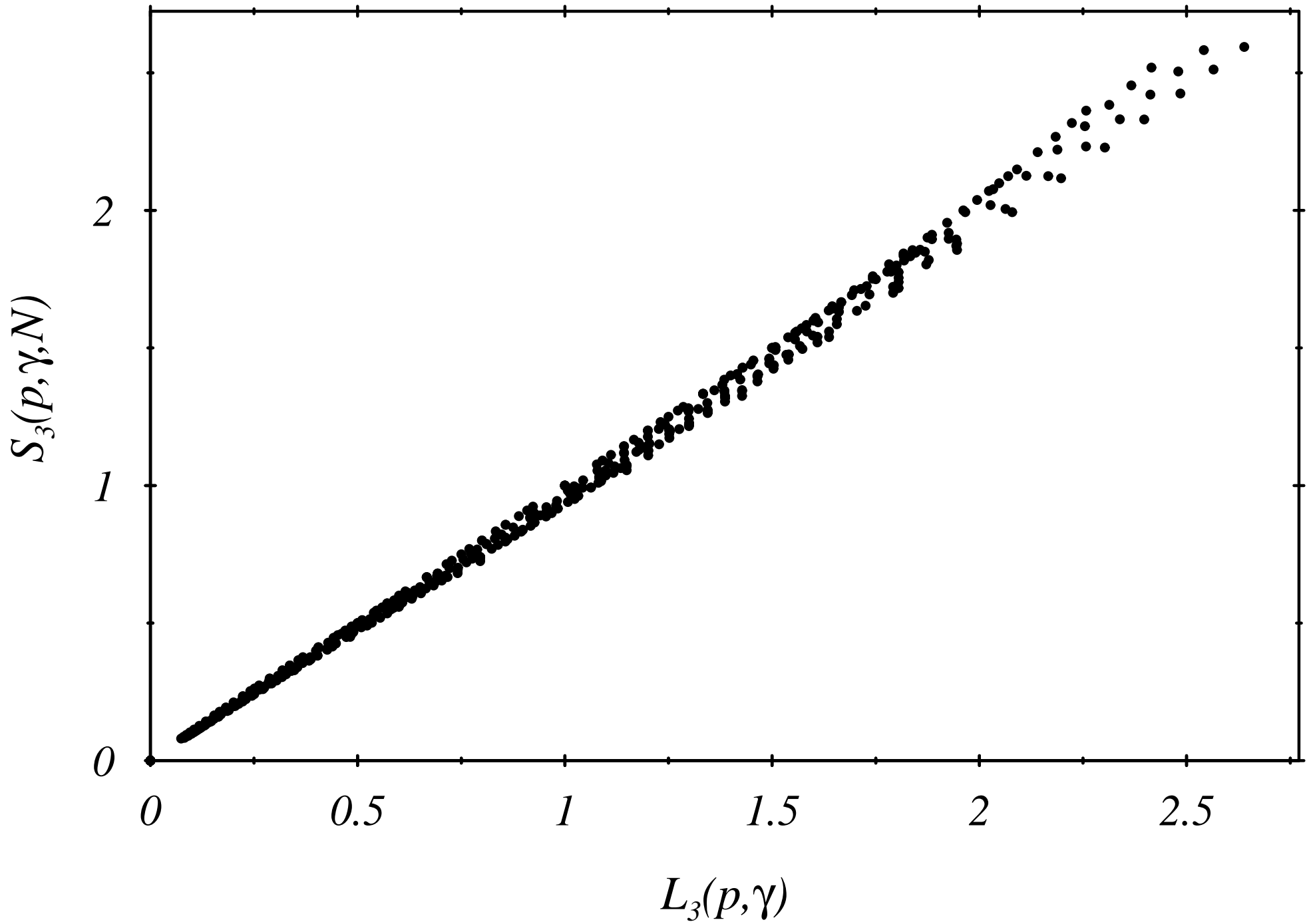
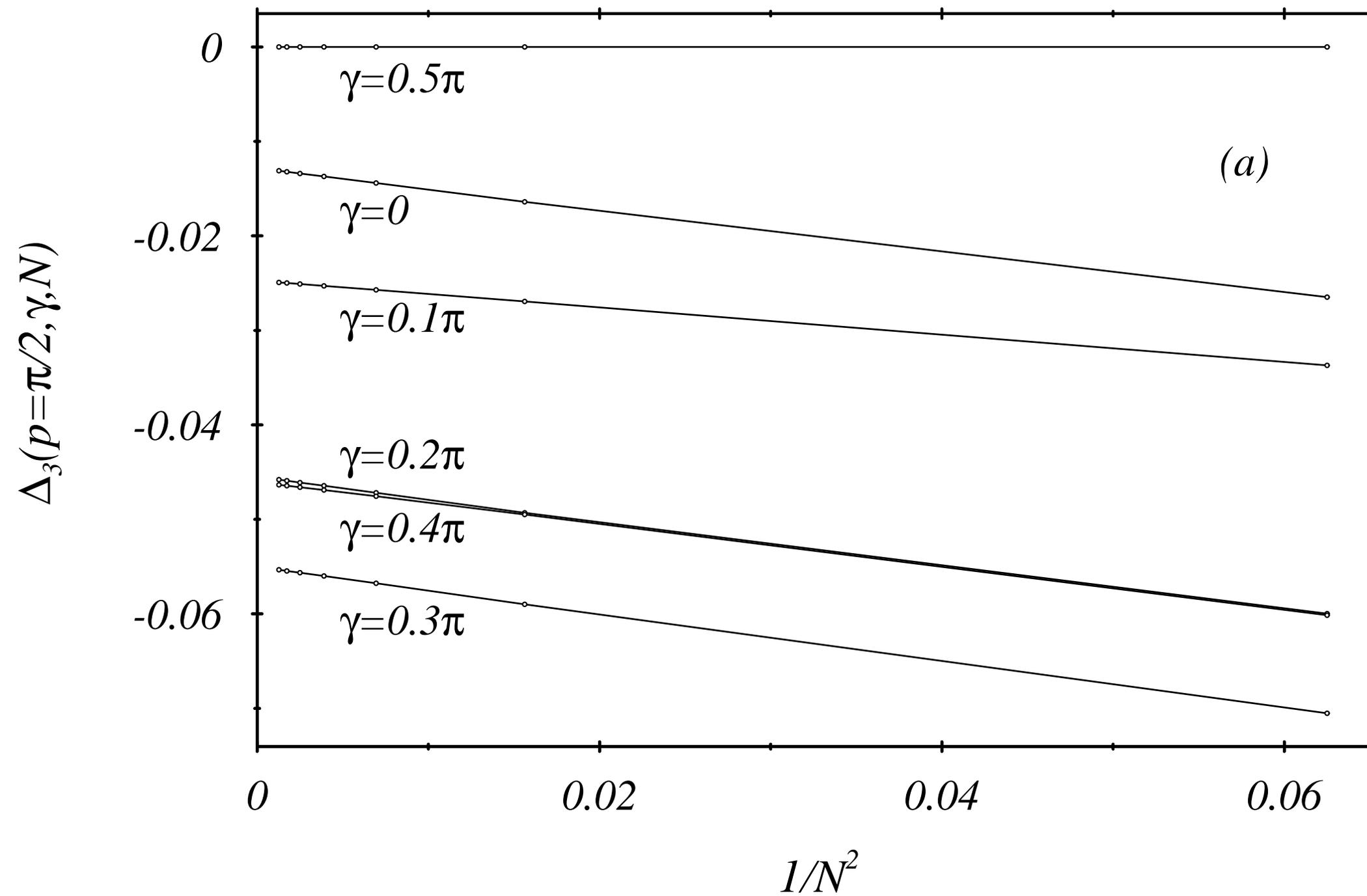
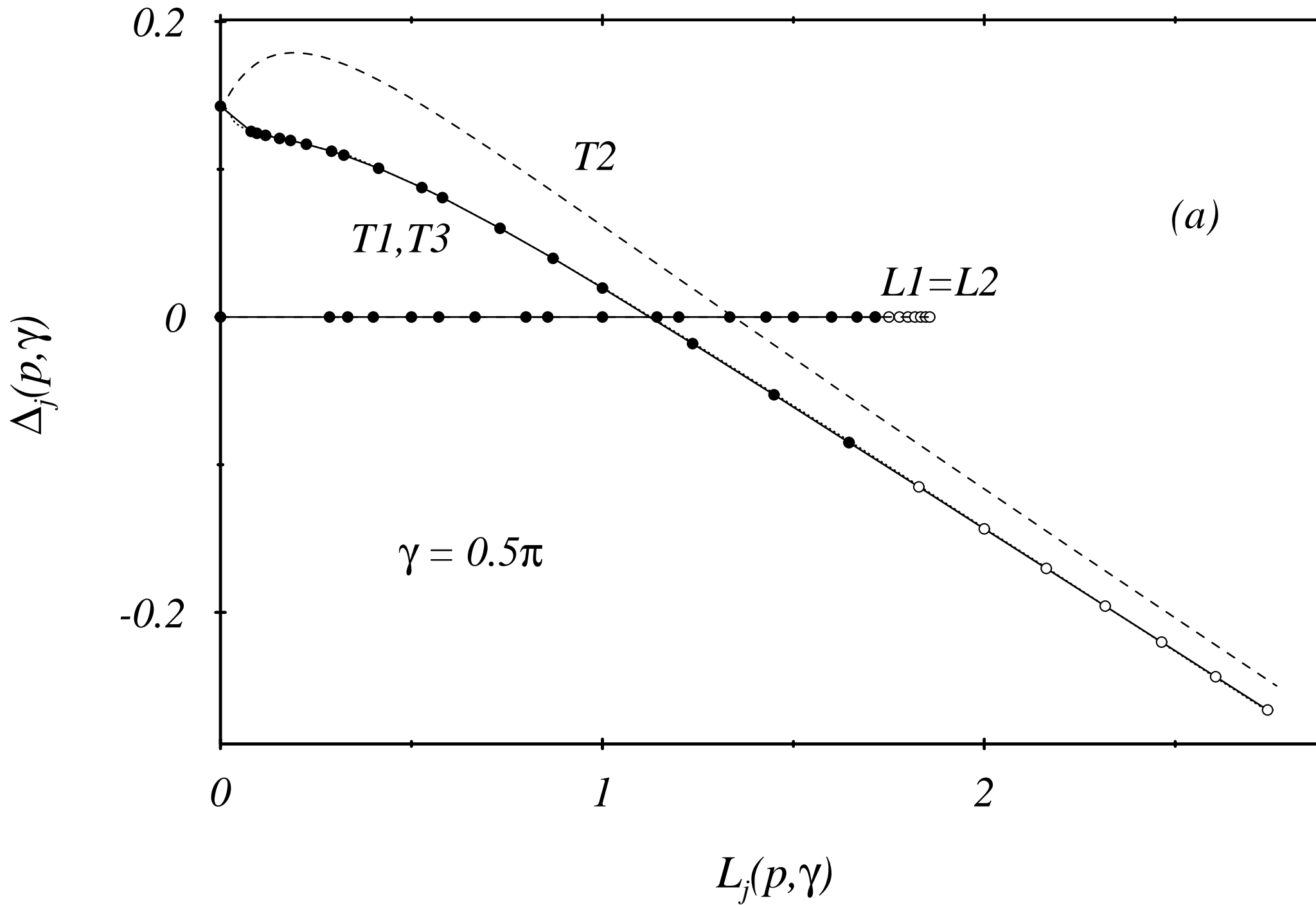
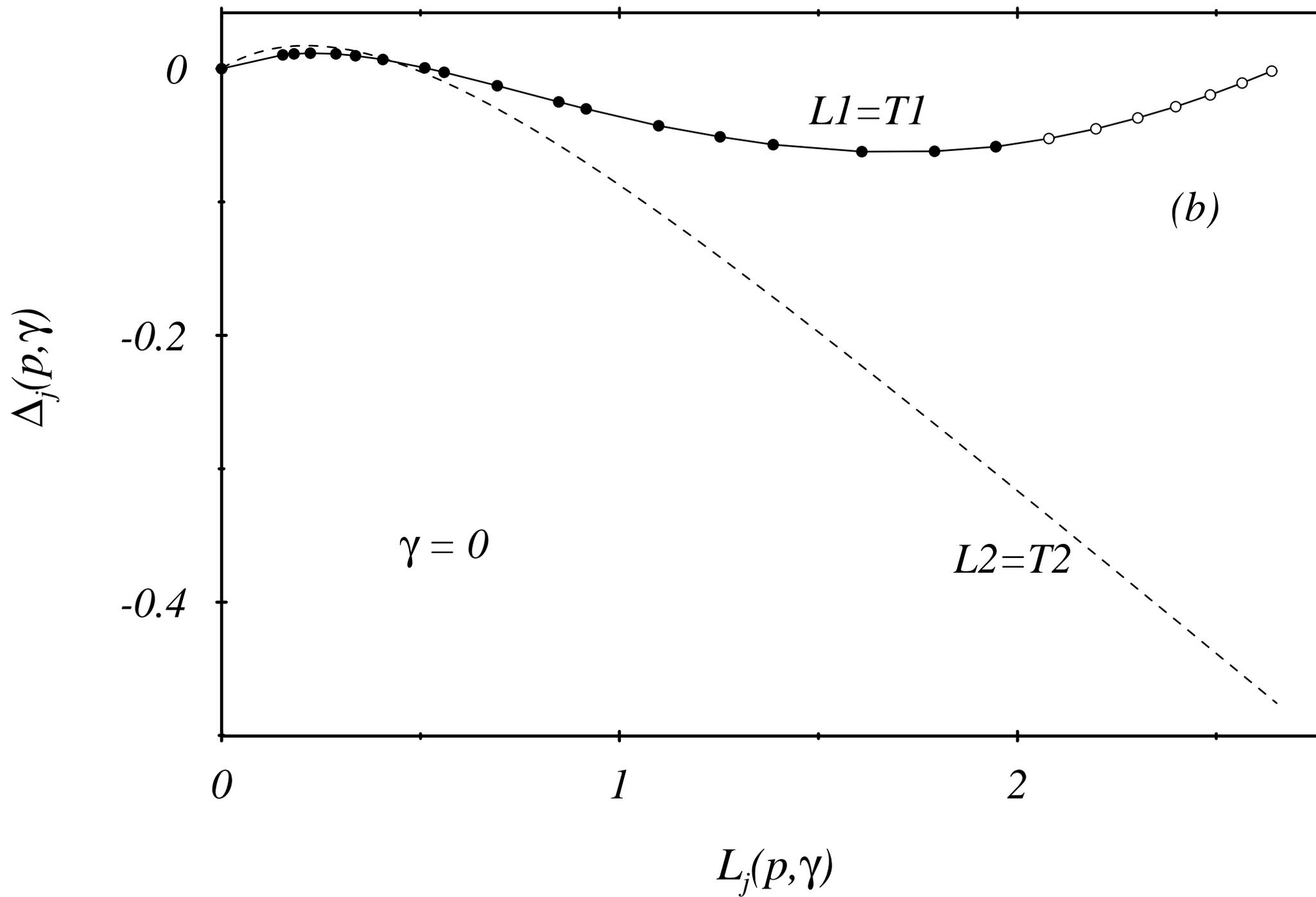
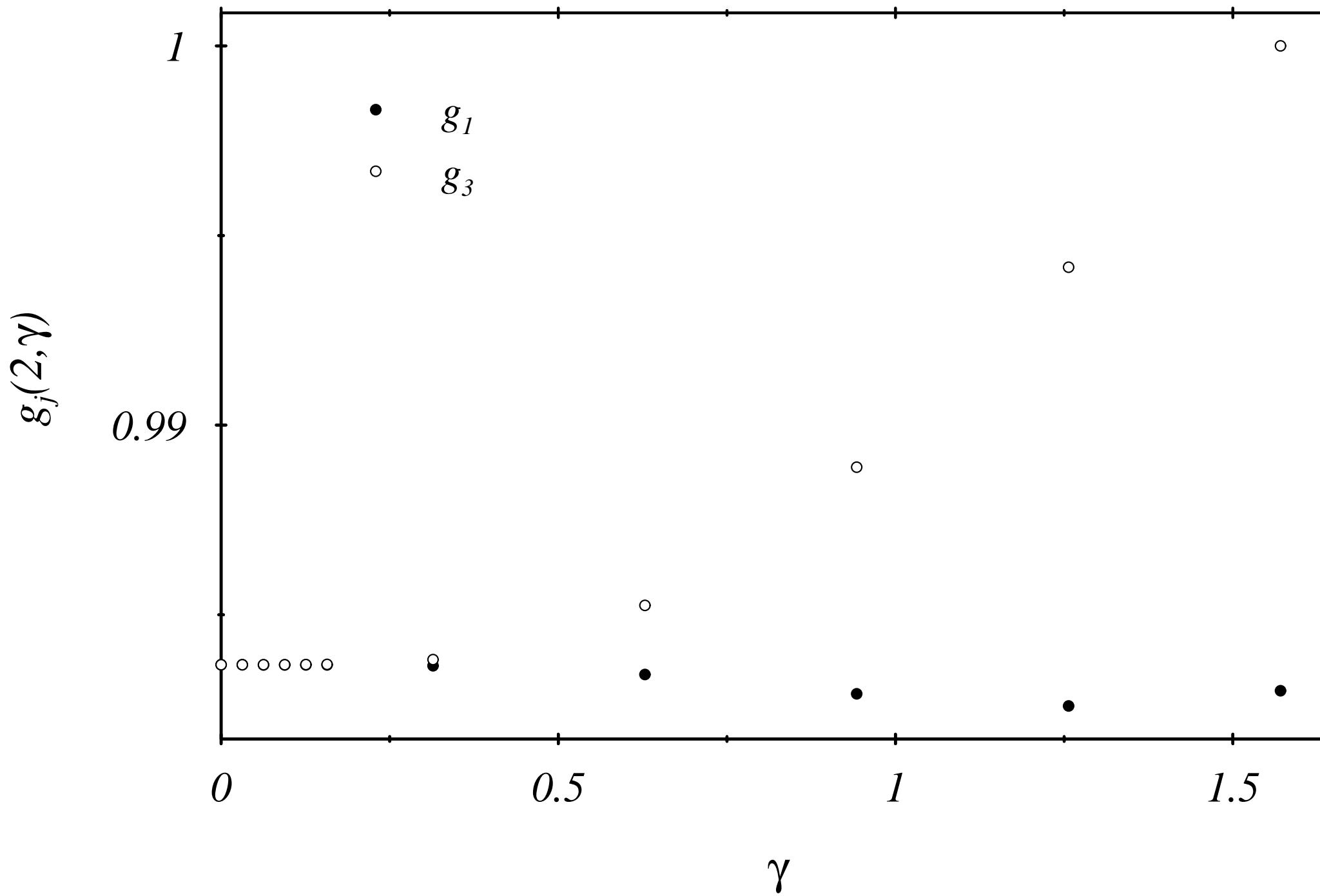
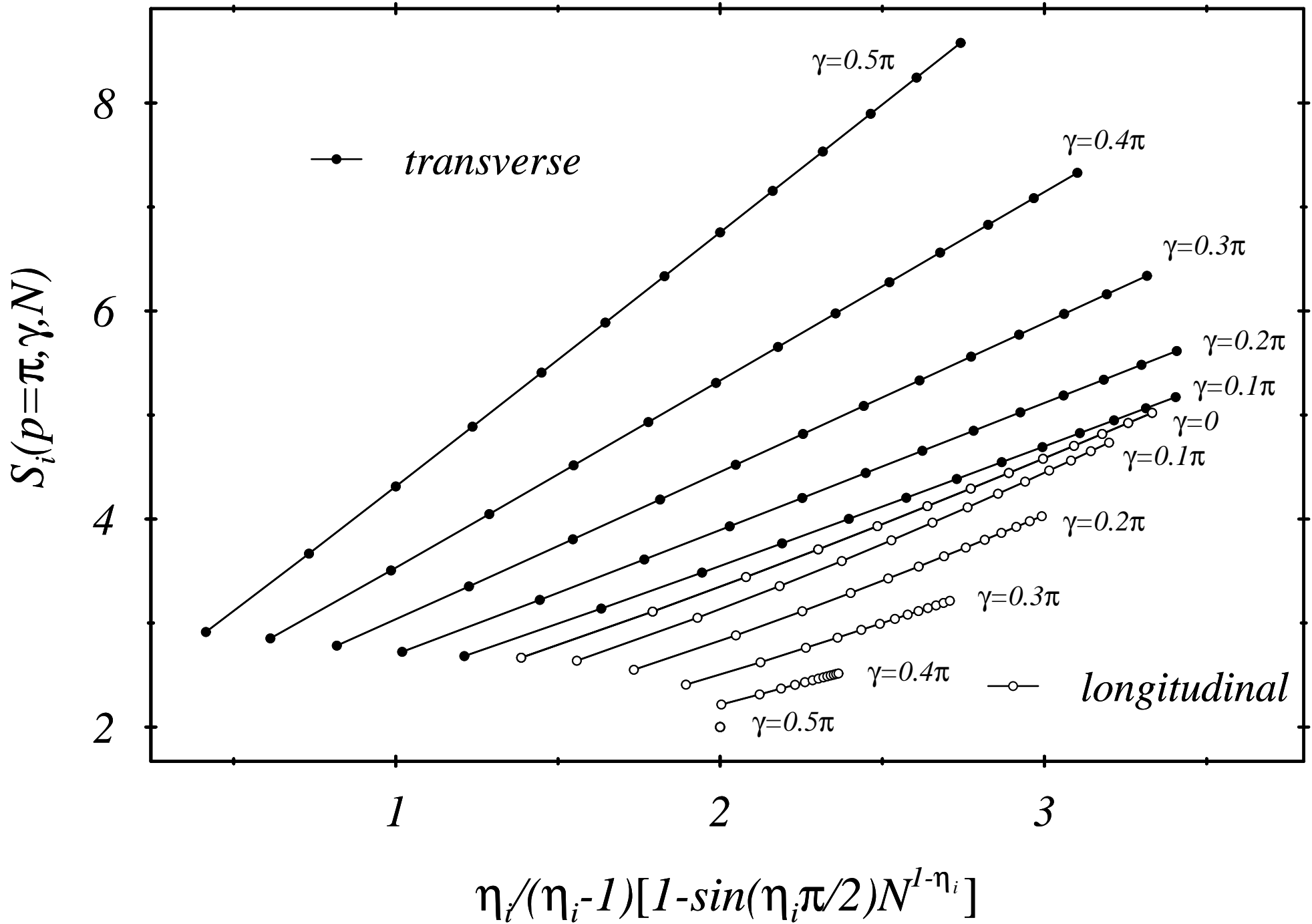
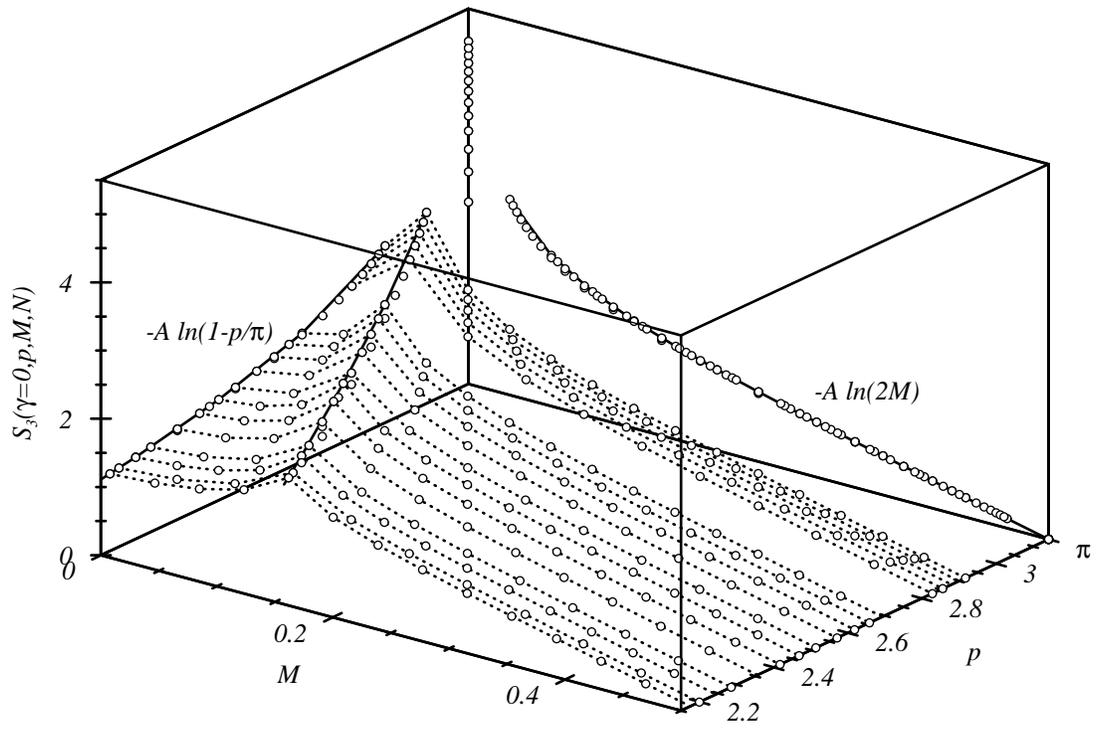
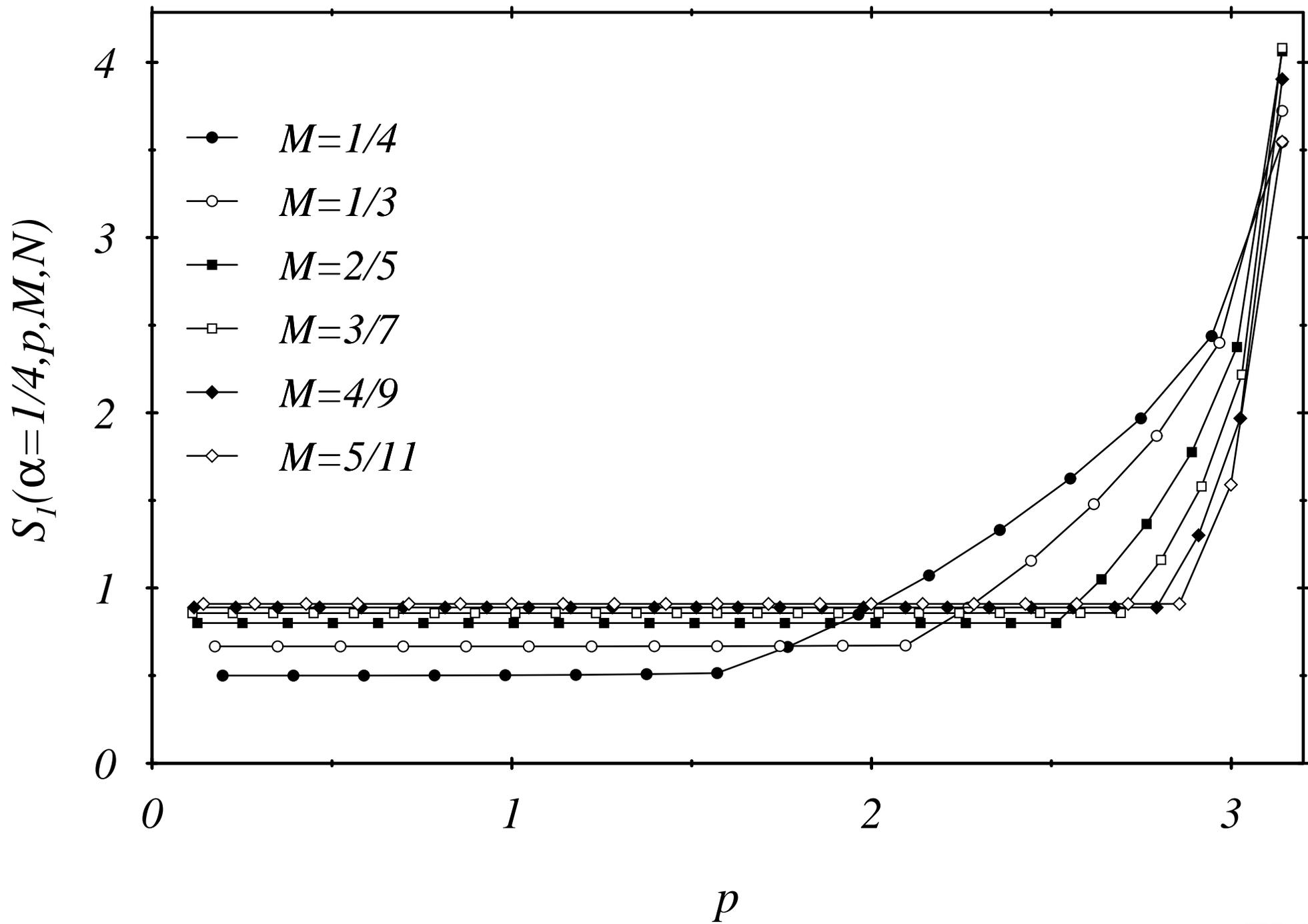
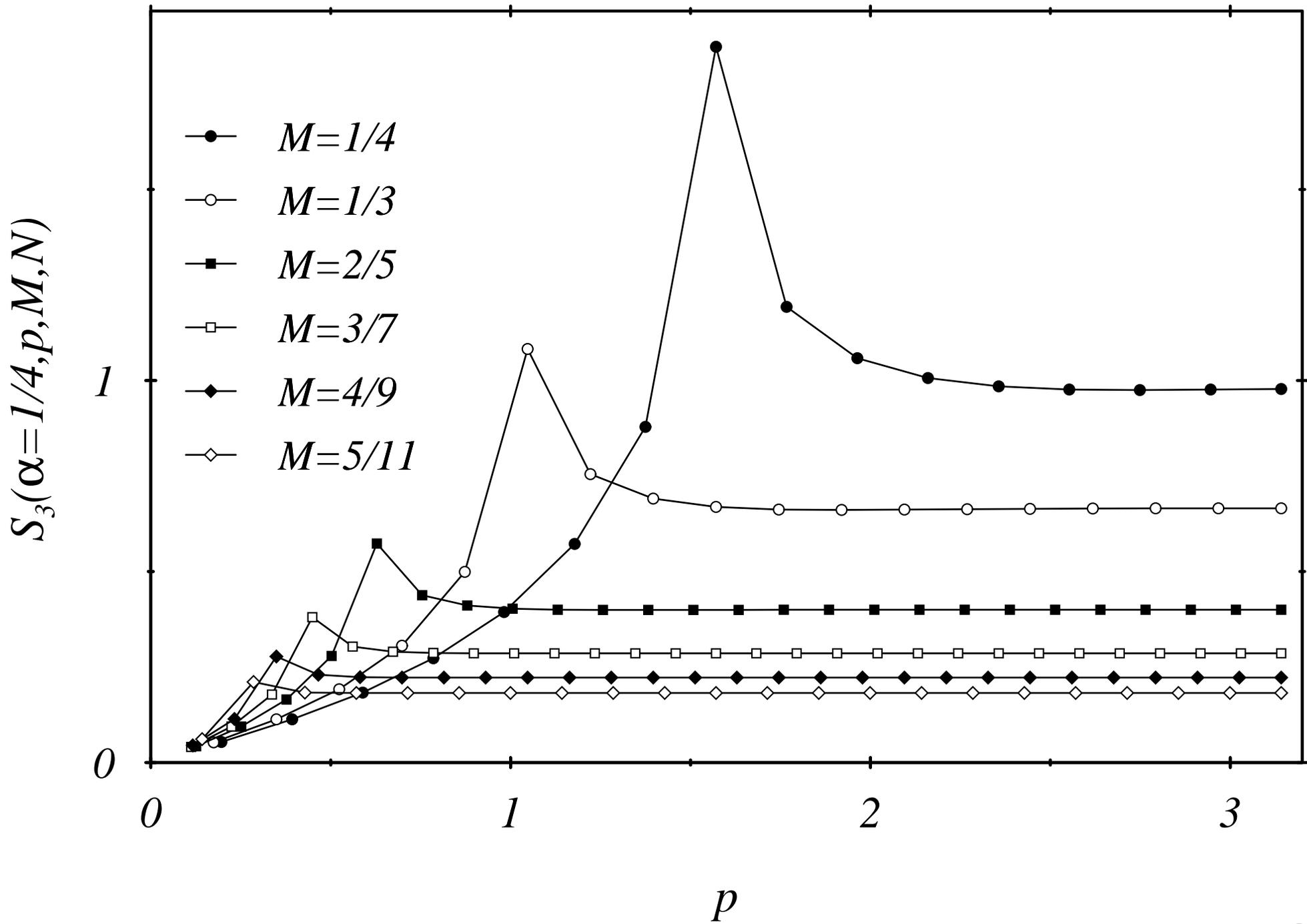
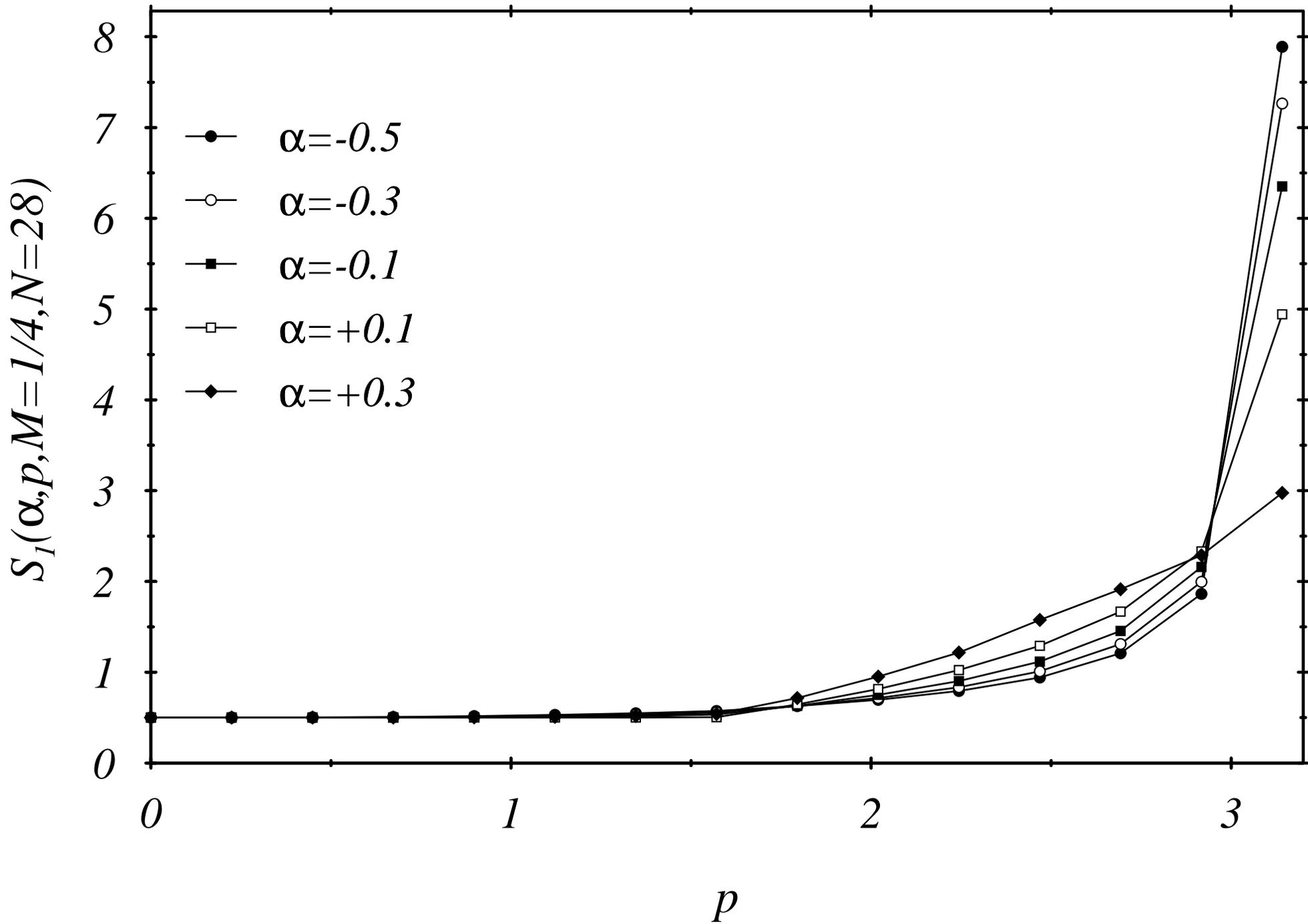
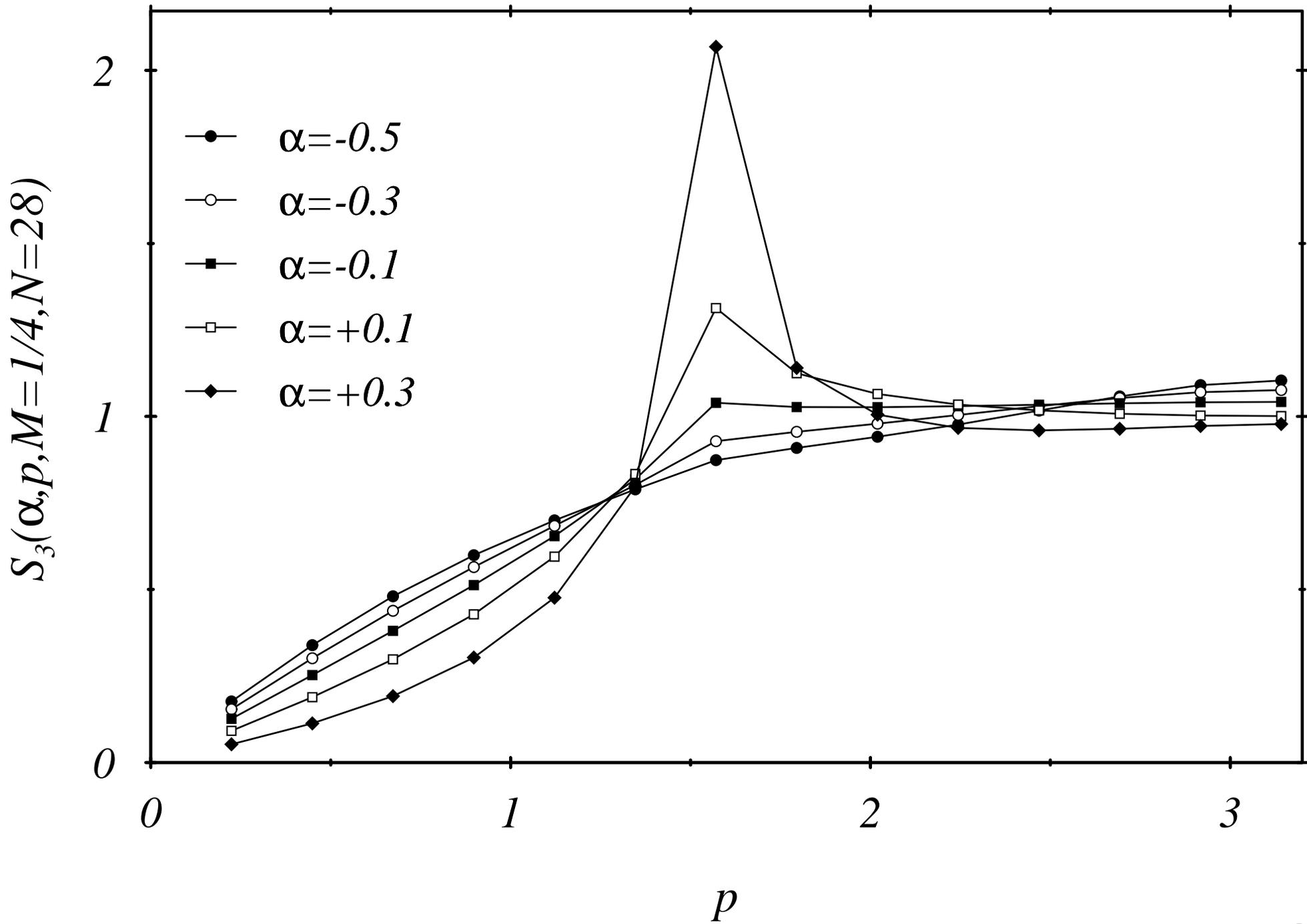
\end{document}